\documentclass[prd,showkeys,floatfix,twocolumn,amsmath,amssymb,floatfix]{revtex4}
\usepackage{graphicx}
\usepackage{multirow}
\usepackage{subfigure}
\usepackage[colorlinks,citecolor=blue,anchorcolor=red,menucolor=red,linkcolor=red,filecolor=red,runcolor=red,urlcolor=blue,frenchlinks=red]{hyperref}
\usepackage{enumitem}
\usepackage{epstopdf}

\makeatletter
\renewcommand{\@thesubfigure}{\hskip\subfiglabelskip}
\makeatother
\allowdisplaybreaks[3]

\begin{document}
\title{Light single-gluon hybrid states with various (exotic) quantum numbers}
%

\author{Wei-Han Tan$^1$}
\author{Niu Su$^{1,2}$}
\author{Hua-Xing Chen$^1$}
\email{hxchen@seu.edu.cn}

\affiliation{$^1$School of Physics, Southeast University, Nanjing 210094, China\\
$^2$Research Center for Nuclear Physics (RCNP), Osaka University, Ibaraki 567-0047, Japan}
\begin{abstract}
We apply the QCD sum rule method to study the light single-gluon hybrid states with various (exotic) quantum numbers. We construct twenty-four single-gluon hybrid currents, and use eighteen of them to calculate the masses of forty-four single-gluon hybrid states with the quark-gluon contents $\bar q q g$ ($q=u/d$) and $\bar s s g$. We concentrate on the hybrid states with the exotic quantum number $J^{PC} = 1^{-+}$, whose masses and widths are calculated to be $M_{|\bar q q g;1^-1^{-+}\rangle} =1.67^{+0.15}_{-0.17}$~GeV, $\Gamma_{|\bar q q g;1^-1^{-+}\rangle} = 530^{+540}_{-330}$~MeV, $M_{|\bar q q g;0^+1^{-+}\rangle} = 1.67^{+0.15}_{-0.17}$~GeV, $\Gamma_{|\bar q q g;0^+1^{-+}\rangle} = 120^{+160}_{-110}$~MeV, $M_{|\bar s s g;0^+1^{-+}\rangle} = 1.84^{+0.14}_{-0.15}$~GeV, and $\Gamma_{|\bar s s g;0^+1^{-+}\rangle} = 100^{+110}_{-~80}$~MeV. Our results support the interpretations of the $\pi_1(1600)$ and $\eta_1(1855)$ as the hybrid states $|\bar q q g;1^-1^{-+}\rangle$ and $|\bar s s g;0^+1^{-+}\rangle$, respectively. Considering the uncertainties, our results suggest that the $\pi_1(1600)$ and $\eta_1(1855)$ may also be interpreted as the hybrid states $|\bar q q g;1^-1^{-+}\rangle$ and $|\bar q q g;0^+1^{-+}\rangle$, respectively. To differentiate these two assignments and to verify whether they are hybrid states or not, we propose to examine the $a_1(1260) \pi$ decay channel in future experiments.
\end{abstract}
\maketitle
\pagenumbering{arabic}

\section{Introduction}
\label{sec:intro}

A single-gluon hybrid state is composed of one valence quark and one valence antiquark as well as one valence gluon. Especially, the hybrid states with $J^{PC} = 0^{\pm-}/1^{-+}/2^{+-}/3^{-+}/\cdots$ are of particular interests, since these exotic quantum numbers can not be accessed by the conventional $\bar q q$ meson~\cite{pdg,Klempt:2007cp,Amsler:2004ps,Bugg:2004xu,Meyer:2010ku,Meyer:2015eta,Chen:2016qju,Briceno:2017max,Ketzer:2019wmd,Jin:2021vct,Chen:2022asf}. Experimental confirmation of the hybrid state is a direct test of QCD in the low energy sector. Recently, the BESIII collaboration performed a partial wave analysis of the $J/\psi \to \gamma \eta \eta^\prime$ decay, and observed the $\eta_1(1855)$ in the $\eta \eta^\prime$ mass spectrum with a statistical significance larger than $19\sigma$~\cite{BESIII:2022riz,BESIII:2022iwi}. This state has the exotic quantum number $I^GJ^{PC} = 0^+1^{-+}$. Its mass and width were measured to be
\begin{eqnarray}
\eta_1(1855) &:& M = 1855 \pm 9 ^{+6}_{-1} {\rm~MeV}/c^2 \, ,
\\ \nonumber && \Gamma = 188 \pm 18 ^{+3}_{-8} {\rm~MeV} \, .
\end{eqnarray}
Besides the isoscalar state $\eta_1(1855)$, there are three isovector states, $\pi_1(1400)$~\cite{IHEP-Brussels-LosAlamos-AnnecyLAPP:1988iqi}, $\pi_1(1600)$~\cite{E852:1998mbq}, and $\pi_1(2015)$~\cite{E852:2004gpn}, which have the exotic quantum number $I^GJ^{PC} = 1^-1^{-+}$. According to PDG, their masses and widths are~\cite{pdg}:
\begin{eqnarray}
\pi_1(1400) &:& M = 1354 \pm 25 {\rm~MeV} \, ,
\\ \nonumber && \Gamma = 330 \pm 35 {\rm~MeV} \, ;
\\ \pi_1(1600) &:& M = 1661 ^{+15}_{-11} {\rm~MeV} \, ,
\\ \nonumber && \Gamma = 240 \pm 50 {\rm~MeV} \, ;
\\ \pi_1(2015) &:& M = 2014\pm20\pm16 {\rm~MeV} \, ,
\\ \nonumber && \Gamma = 230 \pm 32 \pm 73 {\rm~MeV} \, .
\end{eqnarray}
The $\pi_1(1400)$ was observed in the $\eta \pi$ decay channel by several collaborations~\cite{IHEP-Brussels-LosAlamos-AnnecyLAPP:1988iqi,Aoyagi:1993kn,E852:1997gvf,VES:2001rwn,CrystalBarrel:1998cfz,CrystalBarrel:2019zqh}. It was also observed in the $\rho \pi$ decay channel by OBELIX~\cite{OBELIX:2004oio}, but this was not confirmed by COMPASS~\cite{COMPASS:2018uzl}. The $\pi_1(1600)$ was observed in the $\rho \pi$, $\eta^\prime \pi$, $b_1(1235) \pi$, and $f_1(1285) \pi$ decay channels by several collaborations~\cite{E852:1998mbq,Khokhlov:2000tk,Baker:2003jh,COMPASS:2009xrl,CLEO:2011upl}, while the $\pi_1(2015)$ was observed in the $b_1(1235) \pi$ and $f_1(1285) \pi$ decay channels only in the BNL E852 experiments~\cite{E852:2004gpn,E852:2004rfa}. In recent years the COMPASS and JPAC collaborations further examined the $\eta \pi$ and $\eta^\prime \pi$ decay channels~\cite{COMPASS:2014vkj,COMPASS:2018uzl,JPAC:2018zyd,COMPASS:2021ogp}, and their results suggest that there is only one exotic $\pi_1$ resonance coupling to both the $\eta \pi$ and $\eta^\prime \pi$ channels, while there is no evidence for a second exotic resonance. Its mass and width were determined to be $1564\pm24\pm86$~MeV and $492\pm54\pm102$~MeV, respectively~\cite{COMPASS:2014vkj}.

In the past fifty years there have been a lot of theoretical investigations on the hybrid states, such as the MIT bag model~\cite{Barnes:1977hg,Hasenfratz:1980jv,Chanowitz:1982qj}, flux-tube model~\cite{Isgur:1983wj,Close:1994hc,Page:1998gz,Qiu:2022ktc}, AdS/QCD model~\cite{Andreev:2012hw,Bellantuono:2014lra}, lattice QCD~\cite{Michael:1985ne,Juge:2002br,Lacock:1996ny,MILC:1997usn,Dudek:2013yja}, QCD sum
rules~\cite{Reinders:1982hd,Balitsky:1982ps,Govaerts:1983ka,Kisslinger:1995yw,Jin:2002rw,Li:2021fwk}, and constituent gluon model~\cite{Horn:1977rq,Szczepaniak:2001rg,Guo:2007sm}, etc. However, their nature remains elusive since we still poorly understand the gluon degree of freedom. Experimentally, it is not easy to identify the hybrid states unambiguously, so there is currently no definite experimental evidence on their existence. Theoretically, it is also not easy to define the gluon degree of freedom, and a precise definition of the constituent gluon is still lacking. We refer to Refs.~\cite{Horn:1977rq,Szczepaniak:2001rg,Guo:2007sm,Coyne:1980zd,Chanowitz:1980gu,Barnes:1981ac,Cornwall:1982zn,Cho:2015rsa} for discussions on how to construct glueballs and hybrid states using the constituent gluons.

There have been many theoretical calculations on the masses of the $J^{PC} = 1^{-+}$ hybrid states~\cite{Ebert:2009ub,Kim:2008qh,Ping:2009zza,Kitazoe:1983xx,Dudek:2010wm,Dudek:2009qf,Kisslinger:2009pw}. For examples, their masses extracted from the lattice QCD simulations are about 1.7~GeV~\cite{Hedditch:2005zf}, 1.8~GeV~\cite{Bernard:2003jd}, and 2.0~GeV~\cite{McNeile:1998cp}, while their masses extracted from the flux tube model and the constituent gluon model are about 1.9~GeV~\cite{Isgur:1984bm,Burns:2006wz,Iddir:2007dq}. The QCD sum rule method has been widely applied to study the $J^{PC} = 1^{-+}$ hybrid states in Refs.~\cite{Govaerts:1984bk,Chetyrkin:2000tj,Huang:1998zj,Latorre:1985tg,Yang:2007cc,Reinders:1981ww,Narison:2009vj}, and this method has also been applied to study the $J^{PC} = 0^{+-}$ and $2^{+-}$ hybrid states in Refs.~\cite{Ho:2018cat,Wang:2023whb}. We refer to Refs.~\cite{Page:1996rj,Isgur:1985vy,DeViron:1984svx,Frere:1988ac,Zhu:1998sv,Zhu:1999wg,Godfrey:2002rp,McNeile:2006bz,Zhang:2002id,Close:1994pr,Afanasev:1997fp,Szczepaniak:2001qz,Shastry:2022mhk,Chen:2022isv,Yu:2022lwl,Chen:2008ne,Chen:2008qw,Chen:2015moa,Zhang:2019ykd,Tang:2021zti,Dong:2022cuw,Yang:2022rck,Wan:2022xkx,Wang:2022sib,Ji:2022blw,Frere:2024wsf,Barsbay:2024vjt} for more theoretical studies. We also refer to our recent review~\cite{Chen:2022asf} as well as the reviews~\cite{Klempt:2007cp,Amsler:2004ps,Bugg:2004xu,Meyer:2010ku,Meyer:2015eta,Chen:2016qju,Briceno:2017max,Ketzer:2019wmd,Jin:2021vct} for detailed discussions.

Besides the exotic quantum numbers $J^{PC} = 0^{+-}/1^{-+}/2^{+-}$, the light single-gluon hybrid states with other quantum numbers have not been well studied in the literature. Accordingly, in this paper we shall systematically investigate the single-gluon hybrid states with various (exotic) quantum numbers through the QCD sum rule method. Especially, we shall concentrate on the hybrid states with the exotic quantum number $J^{PC}=1^{-+}$ and update the previous calculations on their mass spectrum~\cite{Latorre:1985tg}, given that some QCD parameters have been significantly changed in recent years. We shall also update our previous calculations on their decay properties~\cite{Chen:2010ic,Huang:2010dc,Chen:2022qpd}, with more decay channels taken into account (see the caption of Table~\ref{tab:decay}). Assuming their quark-gluon contents to be either $\bar q q g$ ($q=u/d$) or $\bar s s g$ and their isospin to be either $I=1$ or $I=0$, we shall calculate their masses and widths to be:
\begin{eqnarray}
\nonumber M_{|\bar q q g;1^-1^{-+}\rangle} &=&1.67^{+0.15}_{-0.17}{\rm~GeV} \, ,
\\ \nonumber \Gamma_{|\bar q q g;1^-1^{-+}\rangle} &=& 530^{+540}_{-330}{\rm~MeV} \, ,
\\ \nonumber M_{|\bar q q g;0^+1^{-+}\rangle} &=&1.67^{+0.15}_{-0.17}{\rm~GeV} \, ,
\\ \nonumber \Gamma_{|\bar q q g;0^+1^{-+}\rangle} &=& 120^{+160}_{-110}{\rm~MeV} \, ,
\\ \nonumber M_{|\bar s s g;0^+1^{-+}\rangle} &=& 1.84^{+0.14}_{-0.15}{\rm~GeV} \, ,
\\ \nonumber \Gamma_{|\bar s s g;0^+1^{-+}\rangle} &=& 100^{+110}_{-~80}{\rm~MeV} \, .
\end{eqnarray}

This paper is organized as follows. In Sec.~\ref{sec:current} we construct twenty-four single-gluon hybrid currents with various (exotic) quantum numbers. In Sec.~\ref{sec:sumrule} we use eighteen of them to perform QCD sum rule analyses, and calculate masses of forty-four single-gluon hybrid states with the quark-gluon contents $\bar q q g$ ($q=u/d$) and $\bar s s g$. Based on these results, we systematically study the decay properties of the $J^{PC}=1^{-+}$ hybrid states in Sec.~\ref{sec:decay}. The obtained results are summarized in Sec.~\ref{sec:summary}.

\section{Single-gluon hybrid currents}
\label{sec:current}

In this section we systematically construct the single-gluon hybrid currents using the light quark field $q_a(x)$ and its dual field $\bar q_a(x)$ as well as the gluon field strength tensor $G^n_{\mu\nu}(x)$ and its dual field $\widetilde G^n_{\mu\nu} = G^{n,\rho\sigma} \times \epsilon_{\mu\nu\rho\sigma}/2$, with $a=1\cdots3$ and $n=1\cdots8$ the color indices, and $\mu\cdots\sigma$ the Lorentz indices. Generally speaking, we can construct the single-gluon hybrid currents by combining the color-octet quark-antiquark fields
\begin{eqnarray}
\nonumber &\bar q_a \lambda_n^{ab} \gamma_5 q_b \, , \, \bar q_a \lambda_n^{ab} q_b \, , \, &
\\ & \bar q_a \lambda_n^{ab} \gamma_\mu q_b \, , \, \bar q_a \lambda_n^{ab} \gamma_\mu \gamma_5 q_b \, , \, &
\\ \nonumber & \bar q_a \lambda_n^{ab} \sigma_{\mu\nu} q_b \, , &
\end{eqnarray}
and the color-octet gluon fields
\begin{equation}
G_n^{\alpha\beta} \, , \, \widetilde G_n^{\alpha\beta} \, ,
\end{equation}
together with some Lorentz coefficients $\Gamma^{\mu\nu\cdots\alpha\beta}$.

\begin{figure*}[hbt]
\begin{center}
\includegraphics[width=0.7\textwidth]{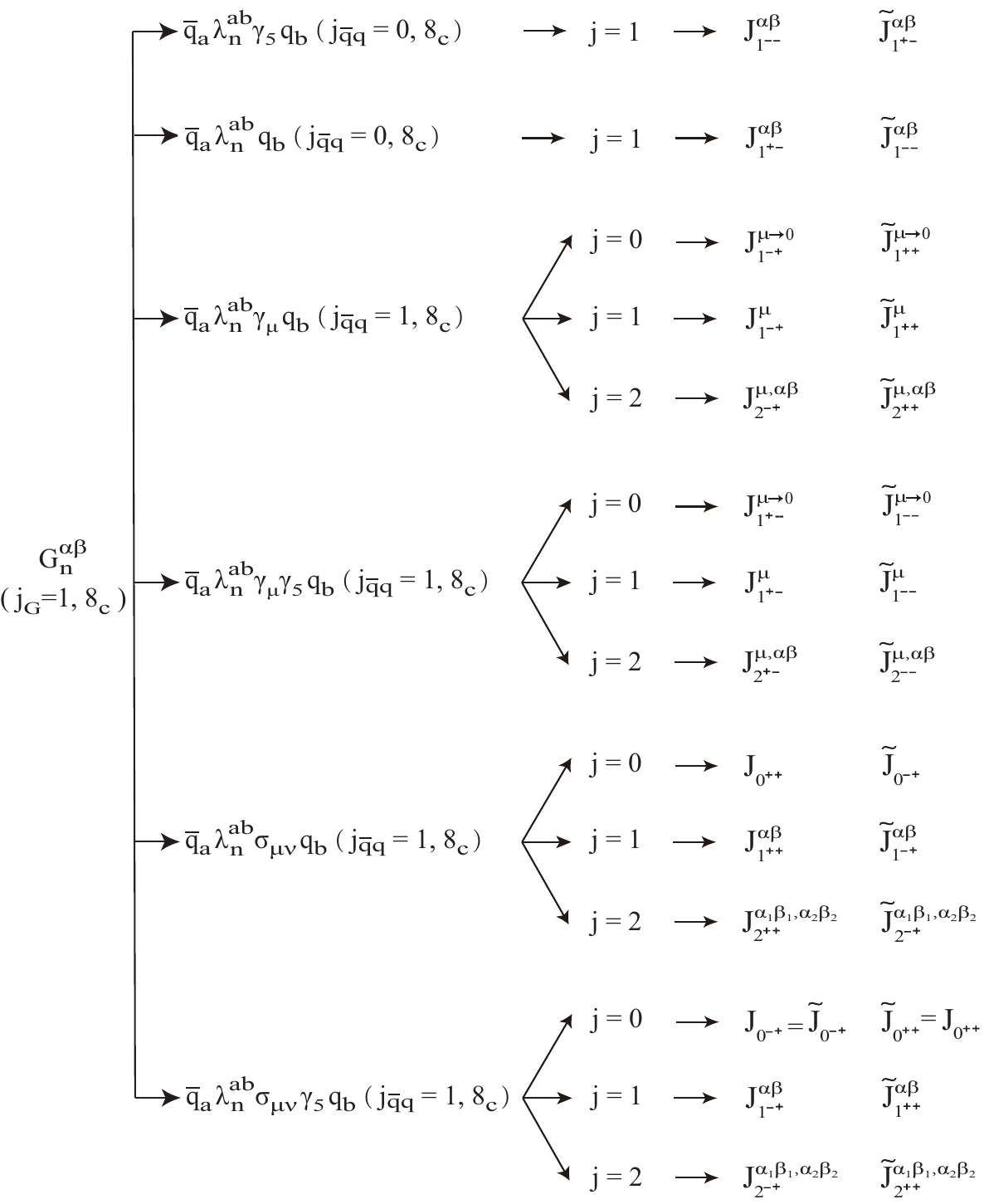}
\caption{Categorization of the single-gluon hybrid currents.}
\label{fig:category}
\end{center}
\end{figure*}

As summarized in Fig.~\ref{fig:category}, there are altogether twenty-four single-gluon hybrid currents, denoted as $J^{\cdots}_{J^{PC}}/\widetilde J^{\cdots}_{J^{PC}}$ with $J$ the total spin:
\begin{eqnarray}
J^{\alpha\beta}_{1^{--}} &=& \bar q_a \lambda_n^{ab} \gamma_5 q_b~g_s G_n^{\alpha\beta} \, ,
\label{def:B1mm}
\\
\widetilde J^{\alpha\beta}_{1^{+-}} &=& \bar q_a \lambda_n^{ab} \gamma_5 q_b~g_s \widetilde G_n^{\alpha\beta} \, ,
\label{def:Bt1pm}
\\
J^{\alpha\beta}_{1^{+-}} &=& \bar q_a \lambda_n^{ab} q_b~g_s G_n^{\alpha\beta} \, ,
\label{def:B1pm}
\\
\widetilde J^{\alpha\beta}_{1^{--}} &=& \bar q_a \lambda_n^{ab} q_b~g_s \widetilde G_n^{\alpha\beta} \, ,
\label{def:Bt1mm}
\\
J^{\mu}_{1^{-+}} &=& \bar q_a \lambda_n^{ab} \gamma_\beta q_b~g_s G_n^{\mu\beta} \, ,
\label{def:A1mp}
\\
\widetilde J^{\mu}_{1^{++}} &=& \bar q_a \lambda_n^{ab} \gamma_\beta q_b~g_s \widetilde G_n^{\mu\beta} \, ,
\label{def:At1pp}
\\
J^{\mu}_{1^{+-}} &=& \bar q_a \lambda_n^{ab} \gamma_\beta \gamma_5 q_b~g_s G_n^{\mu\beta} \, ,
\label{def:A1pm}
\\
\widetilde J^{\mu}_{1^{--}} &=& \bar q_a \lambda_n^{ab} \gamma_\beta \gamma_5 q_b~g_s \widetilde G_n^{\mu\beta} \, ,
\label{def:At1mm}
\\
J^{\mu,\alpha\beta}_{2^{-+}} &=& \bar q_a \lambda_n^{ab} \gamma^\mu q_b~g_s G_n^{\alpha\beta} \, ,
\label{def:A2mp}
\\
\widetilde J^{\mu,\alpha\beta}_{2^{++}} &=& \bar q_a \lambda_n^{ab} \gamma^\mu q_b~g_s \widetilde G_n^{\alpha\beta} \, ,
\label{def:At2pp}
\\
J^{\mu,\alpha\beta}_{2^{+-}} &=& \bar q_a \lambda_n^{ab} \gamma^\mu \gamma_5 q_b~g_s G_n^{\alpha\beta} \, ,
\label{def:A2pm}
\\
\widetilde J^{\mu,\alpha\beta}_{2^{--}} &=& \bar q_a \lambda_n^{ab} \gamma^\mu \gamma_5 q_b~g_s \widetilde G_n^{\alpha\beta} \, ,
\label{def:At2mm}
\\
J_{0^{++}} &=& \bar q_a \lambda_n^{ab} \sigma_{\mu\nu} q_b~g_s G_n^{\mu\nu} \, ,
\label{def:A0pp}
\\
\widetilde J_{0^{-+}} &=& \bar q_a \lambda_n^{ab} \sigma_{\mu\nu} q_b~g_s \widetilde G_n^{\mu\nu} \, ,
\label{def:At0mp}
\\
J_{0^{-+}} &=& \bar q_a \lambda_n^{ab} \sigma_{\mu\nu} \gamma_5 q_b~g_s G_n^{\mu\nu} \, ,
\label{def:A0mp}
\\
\widetilde J_{0^{++}} &=& \bar q_a \lambda_n^{ab} \sigma_{\mu\nu} \gamma_5 q_b~g_s \widetilde G_n^{\mu\nu} \, ,
\label{def:At0pp}
\\
J^{\alpha\beta}_{1^{++}} &=& \mathcal{A}[\bar q_a \lambda_n^{ab} \sigma^{\alpha\mu} q_b~g_s G_{n,\mu}^{\beta}] \, ,
\label{def:B1pp}
\\
\widetilde J^{\alpha\beta}_{1^{-+}} &=& \mathcal{A}[\bar q_a \lambda_n^{ab} \sigma^{\alpha\mu} q_b~g_s \widetilde G_{n,\mu}^{\beta}] \, ,
\label{def:Bt1mp}
\\
J^{\alpha\beta}_{1^{-+}} &=& \mathcal{A}[\bar q_a \lambda_n^{ab} \sigma^{\alpha\mu} \gamma_5 q_b~g_s G_{n,\mu}^{\beta}] \, ,
\label{def:B1mp}
\\
\widetilde J^{\alpha\beta}_{1^{++}} &=& \mathcal{A}[\bar q_a \lambda_n^{ab} \sigma^{\alpha\mu} \gamma_5 q_b~g_s \widetilde G_{n,\mu}^{\beta}] \, ,
\label{def:Bt1pp}
\\
J^{\alpha_1\beta_1,\alpha_2\beta_2}_{2^{++}} &=& \mathcal{S}[\bar q_a \lambda_n^{ab} \sigma^{\alpha_1\beta_1} q_b~g_s G_n^{\alpha_2\beta_2}] \, ,
\label{def:B2pp}
\\
\widetilde J^{\alpha_1\beta_1,\alpha_2\beta_2}_{2^{-+}} &=& \mathcal{S}[\bar q_a \lambda_n^{ab} \sigma^{\alpha_1\beta_1} q_b~g_s \widetilde G_n^{\alpha_2\beta_2}] \, ,
\label{def:Bt2mp}
\\
J^{\alpha_1\beta_1,\alpha_2\beta_2}_{2^{-+}} &=& \mathcal{S}[\bar q_a \lambda_n^{ab} \sigma^{\alpha_1\beta_1} \gamma_5 q_b~g_s G_n^{\alpha_2\beta_2}] \, ,
\label{def:B2mp}
\\
\widetilde J^{\alpha_1\beta_1,\alpha_2\beta_2}_{2^{++}} &=& \mathcal{S}[\bar q_a \lambda_n^{ab} \sigma^{\alpha_1\beta_1} \gamma_5 q_b~g_s \widetilde G_n^{\alpha_2\beta_2}] \, .
\label{def:Bt2pp}
\end{eqnarray}
Especially, we shall concentrate on the fifth current $J^{\mu}_{1^{-+}}$ with the exotic quantum number $J^{PC} = 1^{-+}$, while this current also contains the $J^{PC} = 0^{++}$ component, so we need to separate the $J^{PC} = 1^{-+}$ and $0^{++}$ components in the calculations, as discussed below. In the above expressions $\{\alpha\beta\}/\{\alpha_1\beta_1\}/\{\alpha_2\beta_2\}$ are antisymmetric Lorentz pairs. The four currents $J^{\mu,\alpha\beta}_{2^{-+}}$, $\widetilde J^{\mu,\alpha\beta}_{2^{++}}$, $J^{\mu,\alpha\beta}_{2^{+-}}$, and $\widetilde J^{\mu,\alpha\beta}_{2^{--}}$ all contain three Lorentz indices with the mixed symmetry, so their spin-2 components can not be easily extracted. We shall not investigate these four currents in the present study, and we refer to Ref.~\cite{Wang:2023whb} for detailed discussions. The symbol $\mathcal{A}[\cdots]$ represents anti-symmetrization in the set $\{\alpha\beta\}$, which can be done by multiplying the projection operator
\begin{eqnarray}
\Gamma_{\alpha^\prime\beta^\prime;\alpha\beta} &=& g_{\alpha^\prime \alpha} g_{\beta^\prime \beta} - g_{\beta^\prime \alpha} g_{\alpha^\prime \beta} \, .
\end{eqnarray}
The symbol $\mathcal{S}[\cdots]$ represents symmetrization and subtracting trace terms in the two sets $\{\alpha_1\alpha_2\}$ and $\{\beta_1\beta_2\}$ simultaneously. In the present study we only need to investigate its leading spin-2 component,  which can be done by multiplying the projection operator
\begin{eqnarray}
&& \Gamma^{\prime}_{\alpha_1^\prime\beta_1^\prime,\alpha_2^\prime\beta_2^\prime;\alpha_1\beta_1,\alpha_2\beta_2}
\\ \nonumber &=& (g_{\alpha_1^\prime \alpha_1} g_{\beta_1^\prime \beta_1} - g_{\beta_1^\prime \alpha_1} g_{\alpha_1^\prime \beta_1}) (g_{\alpha_2^\prime \alpha_2} g_{\beta_2^\prime \beta_2} - g_{\beta_2^\prime \alpha_2} g_{\alpha_2^\prime \beta_2})
\\ \nonumber &+& (g_{\alpha_1^\prime \alpha_2} g_{\beta_1^\prime \beta_1} - g_{\beta_1^\prime \alpha_2} g_{\alpha_1^\prime \beta_1}) (g_{\alpha_2^\prime \alpha_1} g_{\beta_2^\prime \beta_2} - g_{\beta_2^\prime \alpha_1} g_{\alpha_2^\prime \beta_2})
\\ \nonumber &+& (g_{\alpha_1^\prime \alpha_1} g_{\beta_1^\prime \beta_2} - g_{\beta_1^\prime \alpha_1} g_{\alpha_1^\prime \beta_2}) (g_{\alpha_2^\prime \alpha_2} g_{\beta_2^\prime \beta_1} - g_{\beta_2^\prime \alpha_2} g_{\alpha_2^\prime \beta_1})
\\ \nonumber &+& (g_{\alpha_1^\prime \alpha_2} g_{\beta_1^\prime \beta_2} - g_{\beta_1^\prime \alpha_2} g_{\alpha_1^\prime \beta_2}) (g_{\alpha_2^\prime \alpha_1} g_{\beta_2^\prime \beta_1} - g_{\beta_2^\prime \alpha_1} g_{\alpha_2^\prime \beta_1})
\\ \nonumber &+& \cdots .
\end{eqnarray}
where $\cdots$ contains the irrelevant terms.

Before performing QCD sum rule analyses, we separately discuss the Lorentz structures of the above single-gluon hybrid currents as follows:
\begin{itemize}

\item Due to the formulae $\sigma_{\mu\nu} \gamma_5 = \epsilon_{\mu\nu\rho\sigma} \sigma^{\rho\sigma} \times i/2$, the two currents $J_{0^{++}}$ and $\widetilde J_{0^{++}}$ are equivalent, and the other two currents $J_{0^{-+}}$ and $\widetilde J_{0^{-+}}$ are also equivalent. Hence, we shall only study the currents $J_{0^{++}}$ and $J_{0^{-+}}$.

\item The current $J^{\alpha\beta}_{1^{--}}$, with two antisymmetric Lorentz indices $\{\alpha\beta\}$, contains both the $J^{PC} = 1^{--}$ and $1^{+-}$ components, so it couples to both the $J^{PC} = 1^{--}$ and $1^{+-}$ states through
\begin{eqnarray}
\langle 0 | J^{\alpha\beta}_{1^{--}} | X_{1^{--}} \rangle &=& i f_{1^{--}} \epsilon^{\alpha\beta \mu \nu} \epsilon_\mu q_\nu \, ,
\label{eq:coupling1}
\\
\langle 0 | J^{\alpha\beta}_{1^{--}} | \widetilde X_{1^{+-}} \rangle &=& i \widetilde f_{1^{+-}} (q^\alpha \epsilon^\beta - q^\beta \epsilon^\alpha) \, ,
\label{eq:coupling2}
\end{eqnarray}
where $f_{1^{--}}$ and $\widetilde f_{1^{+-}}$ are two decay constants. Given the Lorentz structures of Eq.~(\ref{eq:coupling1}) and Eq.~(\ref{eq:coupling2}) to be totally different, we can clearly separate the two states $X_{1^{--}}$ and $\widetilde X_{1^{+-}}$ at the hadron level, {\it i.e.}, we can isolate $X_{1^{--}}$ by investigating the two-point correlation function containing
\begin{eqnarray}
&& \langle 0 | J^{\alpha\beta}_{1^{--}} | X_{1^{--}} \rangle \langle X_{1^{--}} | J^{\alpha^\prime\beta^\prime\dagger}_{1^{--}} | 0 \rangle
\label{eq:coupling3}
\\ \nonumber &=& f_{1^{--}}^2 ~ \epsilon^{\alpha\beta\mu\nu} \epsilon_\mu q_\nu ~ \epsilon^{\alpha^\prime\beta^\prime\mu^\prime\nu^\prime} \epsilon^*_{\mu^\prime} q_{\nu^\prime}
\\ \nonumber &=& - f_{1^{--}}^2 ~ q^2 ~ \left( g^{\alpha \alpha^\prime} g^{\beta \beta^\prime} - g^{\alpha \beta^\prime} g^{\beta \alpha^\prime} \right) + \cdots \, ,
\end{eqnarray}
while the correlation function of $\widetilde X_{1^{+-}}$ dose not contain the above coefficient. It is not so easy to isolate $\widetilde X_{1^{+-}}$ from $J^{\alpha\beta}_{1^{--}}$, but instead, we can study the dual current $\widetilde J^{\alpha\beta}_{1^{+-}}$ that couples to $X_{1^{--}}$ and $\widetilde X_{1^{+-}}$ in the opposite ways. According to the above analysis, we shall study the single-gluon hybrid currents $J^{\alpha\beta}_{1^{--}}/\widetilde J^{\alpha\beta}_{1^{+-}}/J^{\alpha\beta}_{1^{+-}}/\widetilde J^{\alpha\beta}_{1^{--}}/J^{\alpha\beta}_{1^{++}}/\widetilde J^{\alpha\beta}_{1^{-+}}/J^{\alpha\beta}_{1^{-+}}/\widetilde J^{\alpha\beta}_{1^{++}}$ to investigate the single-gluon hybrid states of $J^{PC} = 1^{--}/1^{+-}/1^{+-}/1^{--}/1^{++}/1^{-+}/1^{-+}/1^{++}$, respectively.

\item The four currents $J^{\alpha_1\beta_1,\alpha_2\beta_2}_{2^{++}}$, $\widetilde J^{\alpha_1\beta_1,\alpha_2\beta_2}_{2^{-+}}$, $J^{\alpha_1\beta_1,\alpha_2\beta_2}_{2^{-+}}$, and $\widetilde J^{\alpha_1\beta_1,\alpha_2\beta_2}_{2^{++}}$ all contain various $J^{PC}$ components and so couple to many states. We shall only investigate the four $J=2$ states of $J^{PC} = 2^{++}/2^{-+}/2^{-+}/2^{++}$ by calculating the highest-order correlation functions, {\it e.g.}:
\begin{eqnarray}
&& \Pi_{2^{++}}^{\alpha_1\beta_1,\alpha_2\beta_2;\alpha_1^\prime\beta_1^\prime,\alpha_2^\prime\beta_2^\prime}(q^2)
\\ \nonumber &=& i \int d^4x e^{iqx} \langle 0 | {\bf T}[J^{\alpha_1\beta_1,\alpha_2\beta_2}_{2^{++}}(x) J^{\alpha_1^\prime\beta_1^\prime,\alpha_2^\prime\beta_2^\prime\dagger}_{2^{++}}(0)] | 0 \rangle
\\ \nonumber &=& \mathcal{S}^{\prime} [g^{\alpha_1\alpha_1^\prime}g^{\beta_1\beta_1^\prime}g^{\alpha_2\alpha_2^\prime}g^{\beta_2\beta_2^\prime}]~\Pi_{2^{++}} (q^2) + \cdots \, ,
\end{eqnarray}
where $\mathcal{S}^{\prime}[\cdots]$ represents symmetrization and subtracting trace terms in the four sets $\{\alpha_1\alpha_2\}$, $\{\beta_1\beta_2\}$, $\{\alpha_1^\prime\alpha_2^\prime\}$, and $\{\beta_1^\prime\beta_2^\prime\}$ simultaneously. The correlation function $\Pi_{2^{++}} (q^2)$ is contributed only by the $J^{PC} = 2^{++}$ component, while $\cdots$ contains the contributions from all the $J^{PC}$ components coupling to $J^{\alpha_1\beta_1,\alpha_2\beta_2}_{2^{++}}$. However, we do not know the explicit expression of the rearrangement $\mathcal{S}^{\prime}[\cdots]$, so we are not able to calculate the decay constants of these four currents.

\item The current $J^{\mu}_{1^{-+}}$ contains both the $J^{PC} = 1^{-+}$ and $0^{++}$ components, so it couples to both the $J^{PC} = 1^{+-}$ and $0^{++}$ states through
\begin{eqnarray}
\langle 0 | J^{\mu}_{1^{-+}} | X_{1^{-+}} \rangle &=& \epsilon^\mu f_{1^{-+}} \, ,
\label{eq:coupling4}
\\ \langle 0 | J^{\mu}_{1^{-+}} | X_{0^{++}} \rangle &=& q^\mu f_{0^{++}} \, ,
\label{eq:coupling5}
\end{eqnarray}
where $f_{1^{-+}}$ and $f_{0^{++}}$ are two decay constants. We can clearly separate the two states $X_{1^{-+}}$ and $X_{0^{++}}$ at the hadron level by calculating both $\Pi_{1^{-+}}(q^2)$ and $\Pi_{0^{++}}(q^2)$ of the two-point correlation function
\begin{eqnarray}
&& \Pi_{1^{-+}}^{\mu\nu}(q^2)
\label{eq:coupling6}
\\ \nonumber &\equiv& i \int d^4x e^{iqx} \langle 0 | {\bf T}[J^{\mu}_{1^{-+}}(x) J^{\nu\dagger}_{1^{-+}}(0)] | 0 \rangle
\\ \nonumber &=& (g^{\mu\nu} - q^\mu q^\nu/q^2 )~\Pi_{1^{-+}}(q^2) + ( q^\mu q^\nu/q^2 )~ \Pi_{0^{++}}(q^2) \, .
\end{eqnarray}
Similarly, we shall study the single-gluon hybrid currents $\widetilde J^{\mu}_{1^{++}}/J^{\mu}_{1^{+-}}/\widetilde J^{\mu}_{1^{--}}$ to investigate the single-gluon hybrid states of both $J^{PC} = 1^{++}/1^{+-}/1^{--}$ and $J^{PC} = 0^{-+}/0^{--}/0^{+-}$.

\end{itemize}

%
\section{QCD sum rule analyses}
\label{sec:sumrule}
%

In this section we use the single-gluon hybrid currents given in Eqs.~(\ref{def:B1mm}-\ref{def:At1mm}) and Eqs.~(\ref{def:A0pp}-\ref{def:Bt2pp}) to perform QCD sum rule analyses. We use the current $J^{\mu}_{1^{-+}}$ given in Eq.~(\ref{def:A1mp}) as an example. Based on Eq.~(\ref{eq:coupling4}) and Eq.~(\ref{eq:coupling6}), we study its two-point correlation function
\begin{eqnarray}
\nonumber \Pi_{1^{-+}}^{\mu\nu}(q^2) &=& i \int d^4x e^{iqx} \langle 0 | {\bf T}[J^{\mu}_{1^{-+}}(x) J^{\nu\dagger}_{1^{-+}}(0)] | 0 \rangle
\\ &=& (g^{\mu\nu} - q^\mu q^\nu/q^2 )~\Pi_{1^{-+}}(q^2) + \cdots ,
\label{eq:correlation}
\end{eqnarray}
at both the hadron and quark-gluon levels, where $\Pi_{1^{-+}}(q^2)$ is contributed by the $J^{PC} = 1^{-+}$ state $X_{1^{-+}}$, and $\cdots$ is contributed by the $J^{PC} = 0^{++}$ state $X_{0^{++}}$.

We use the dispersion relation to express $\Pi_{1^{-+}}(q^2)$ as
\begin{equation}
\Pi_{1^{-+}}(q^2) = \int^\infty_{s_<}\frac{\rho_{1^{-+}}(s)}{s-q^2-i\varepsilon}ds \, ,
\label{eq:hadron}
\end{equation}
where $\rho_{1^{-+}}(s) \equiv {\rm Im}\Pi_{1^{-+}}(s)/\pi$ is the spectral density, and $s_< = 4 m_q^2$ is the physical threshold.

At the hadron level, we parameterize $\rho_{1^{-+}}^{\rm phen}(s)$ as one pole dominance for the state $X_{1^{-+}}$ together with a continuum contribution:
\begin{eqnarray}
&& \rho_{1^{-+}}^{\rm phen}(s) \times (g^{\mu\nu} - q^\mu q^\nu/q^2 )
\\ \nonumber &\equiv& \sum_n\delta(s-M^2_n) \langle 0| J^{\mu}_{1^{-+}} | n\rangle \langle n| J^{\nu\dagger}_{1^{-+}} |0 \rangle
\\ \nonumber &=& f_{1^{-+}}^2 \delta(s-M^2_{1^{-+}}) \times (g^{\mu\nu} - q^\mu q^\nu/q^2 ) + \rm{continuum} \, .
\label{eq:rho}
\end{eqnarray}

At the quark-gluon level, we calculate $\rho_{1^{-+}}^{\rm OPE}(s)$ through the method of operator product expansion (OPE). After performing the Borel transformation at both the hadron and quark-gluon levels, we use $\rho_{1^{-+}}^{\rm OPE}(s)$ above the threshold value $s_0$ to approximate the continuum, and derive the sum rule equation
%
\begin{eqnarray}
\Pi_{1^{-+}}(s_0, M_B^2) &\equiv& f^2_{1^{-+}} e^{-M_{1^{-+}}^2/M_B^2}
\\ \nonumber &=& \int^{s_0}_{s_<} e^{-s/M_B^2}\rho_{1^{-+}}^{\rm OPE}(s)ds ,
\end{eqnarray}
%
which can be used to further derive
%
\begin{eqnarray}
M^2_{1^{-+}}(s_0, M_B) &=& \frac{\int^{s_0}_{s_<} e^{-s/M_B^2}s\rho_{1^{-+}}^{\rm OPE}(s)ds}{\int^{s_0}_{s_<} e^{-s/M_B^2}\rho_{1^{-+}}^{\rm OPE}(s)ds} ,
\label{eq:LSR}
\\
f^2_{1^{-+}}(s_0, M_B) &=& \Pi_{1^{-+}}(s_0, M_B^2) \times e^{M_{1^{-+}}^2/M_B^2} .
\label{eq:decay}
\end{eqnarray}
%

\begin{figure}[hbtp]
\begin{center}
\subfigure[(a)]{
\scalebox{0.12}{\includegraphics{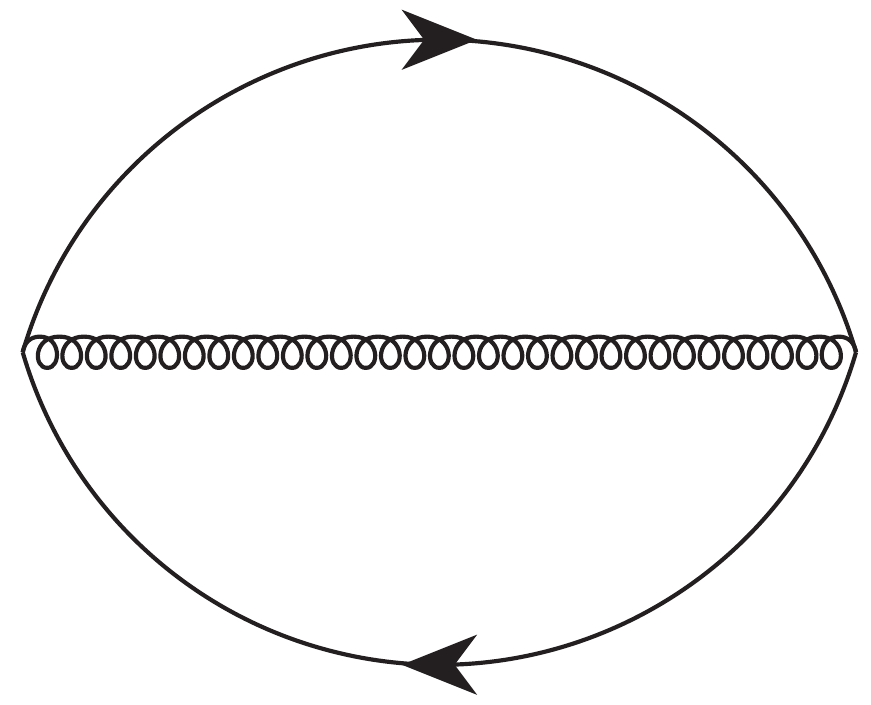}}}
\\
\subfigure[(b--1)]{
\scalebox{0.12}{\includegraphics{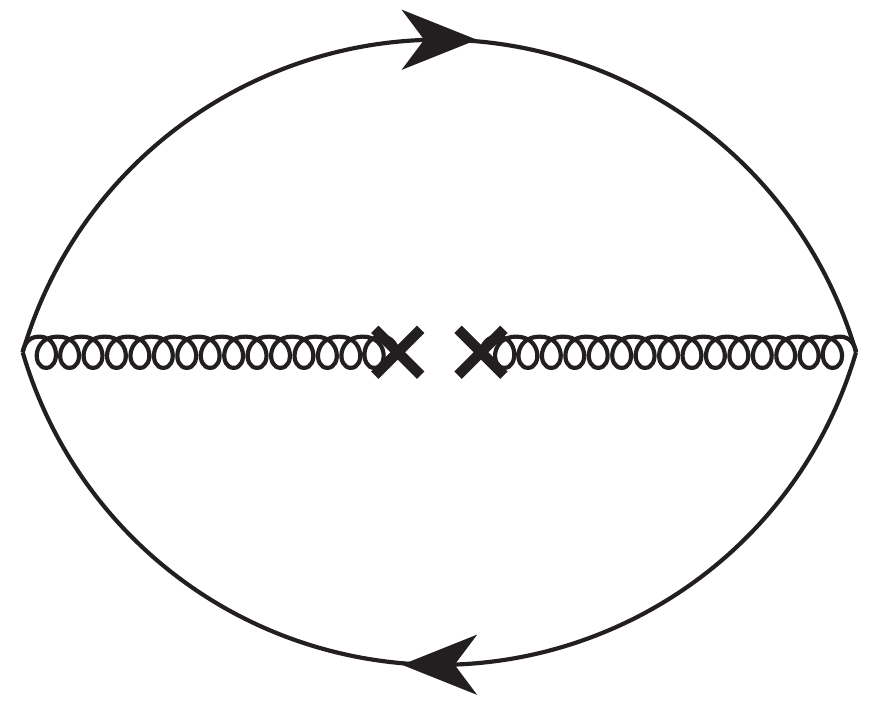}}}~
\subfigure[(b--2)]{
\scalebox{0.12}{\includegraphics{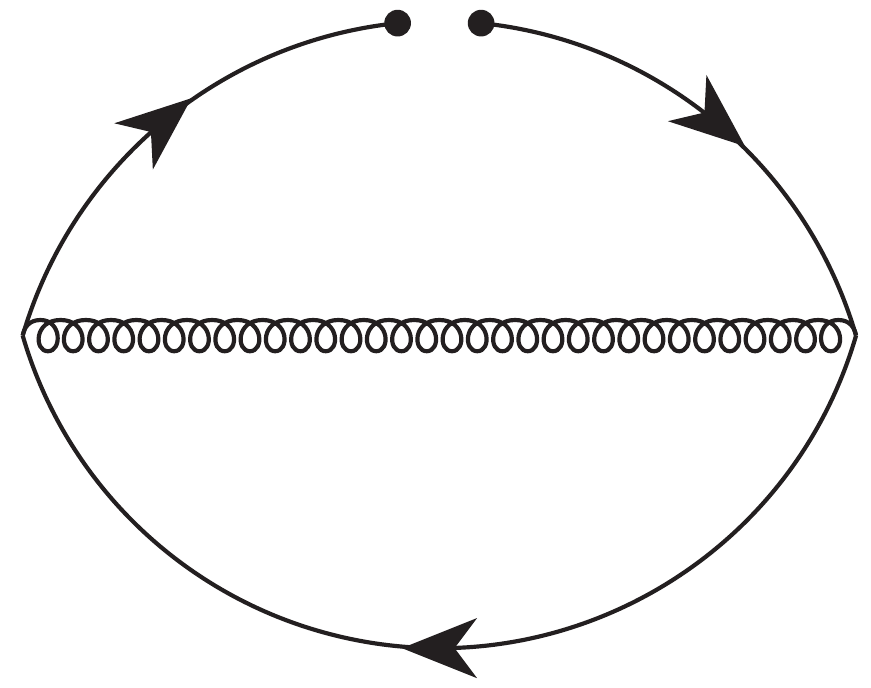}}}~
\subfigure[(b--3)]{
\scalebox{0.12}{\includegraphics{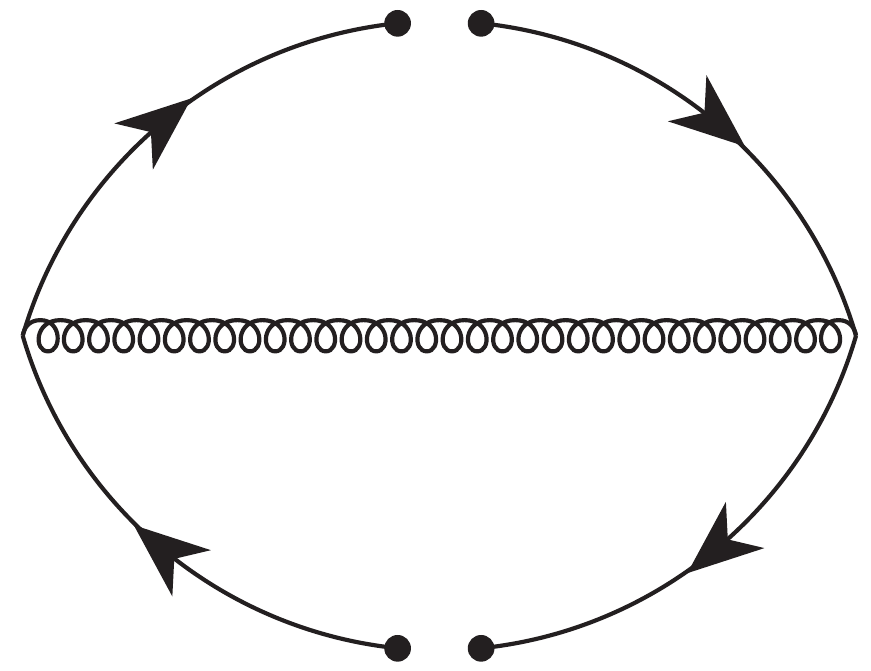}}}
\\
\subfigure[(c--1)]{
\scalebox{0.12}{\includegraphics{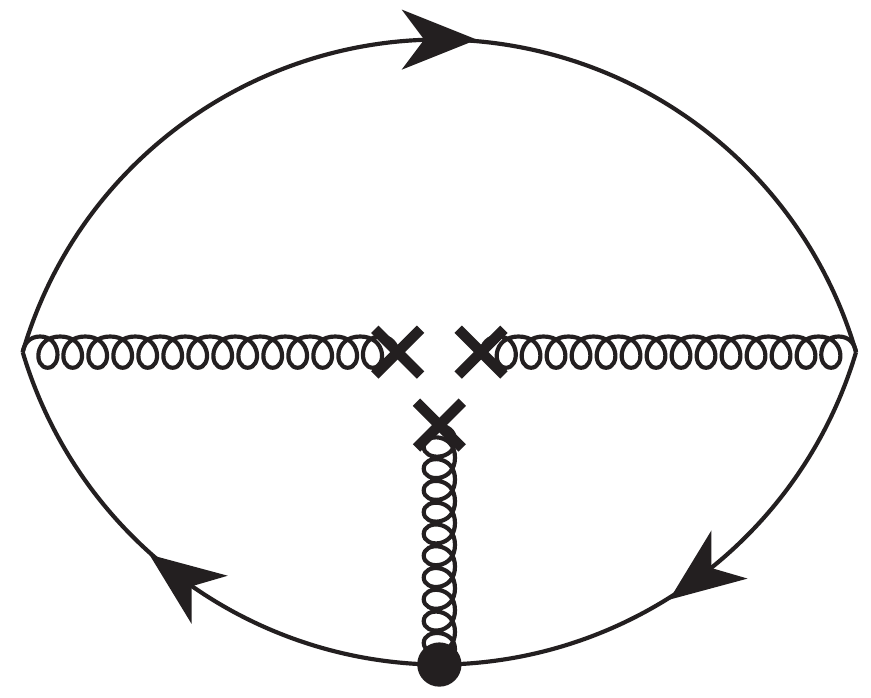}}}~
\subfigure[(c--2)]{
\scalebox{0.12}{\includegraphics{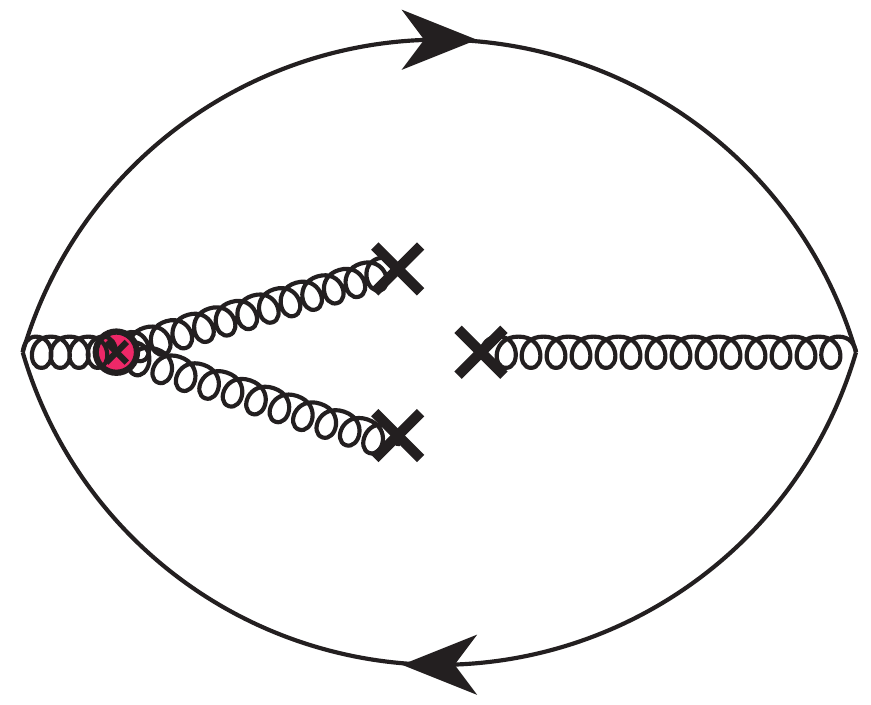}}}~
\\
\subfigure[(d--1)]{
\scalebox{0.12}{\includegraphics{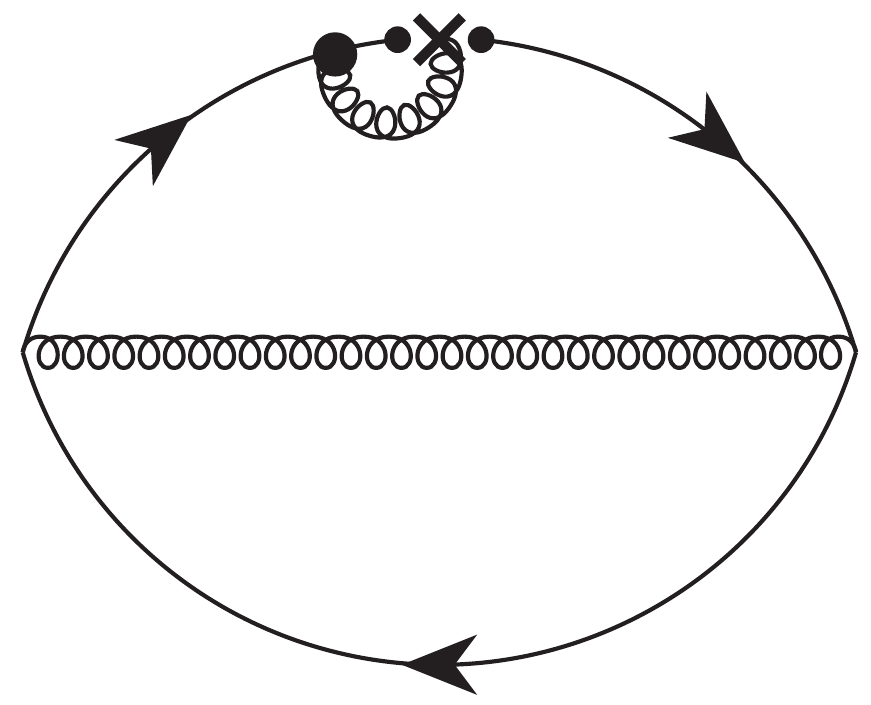}}}~
\subfigure[(d--2)]{
\scalebox{0.12}{\includegraphics{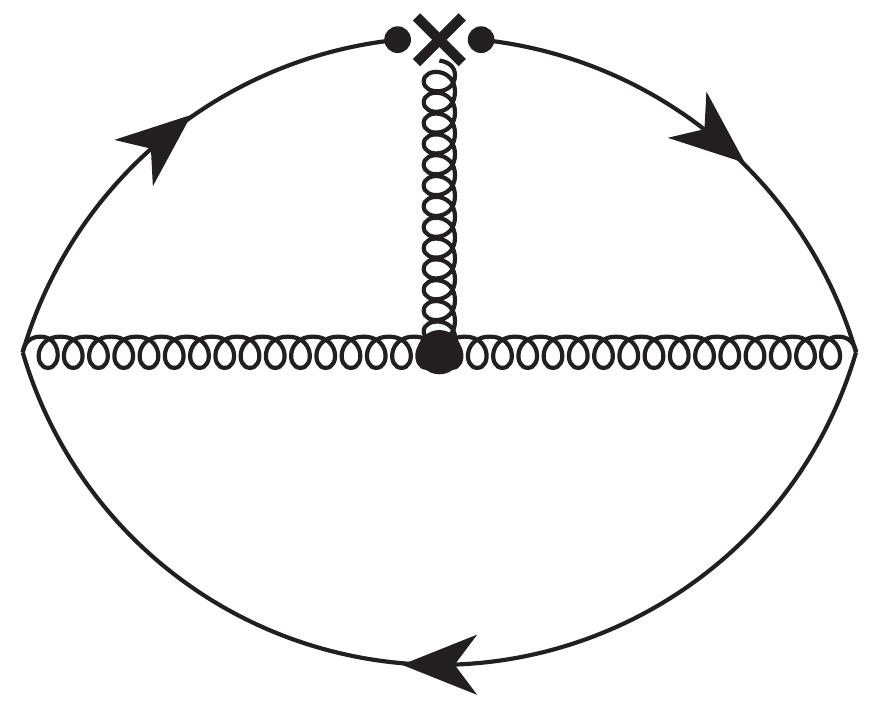}}}~
\subfigure[(d--3)]{
\scalebox{0.12}{\includegraphics{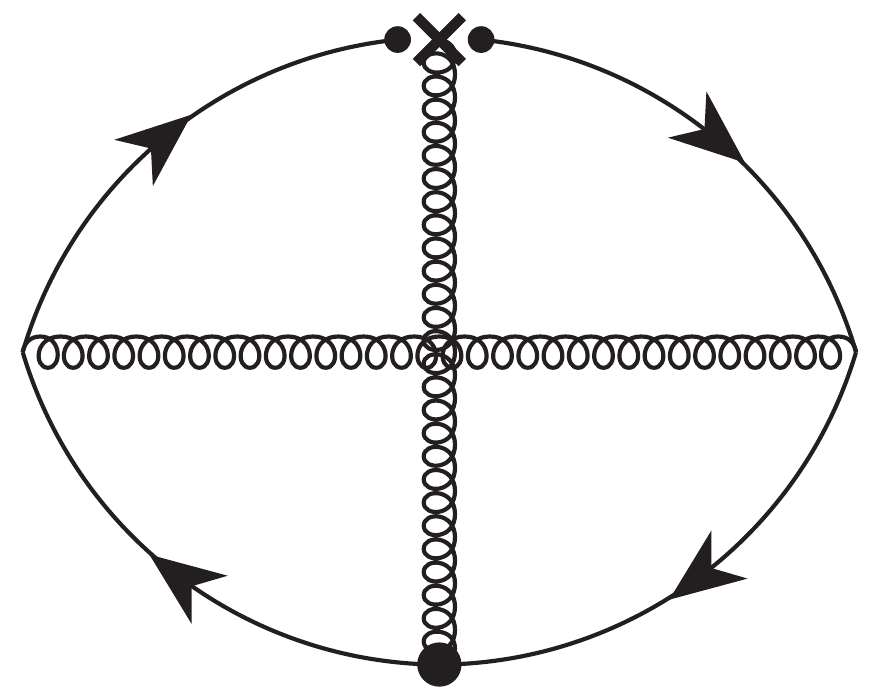}}}
\subfigure[(d--4)]{
\scalebox{0.12}{\includegraphics{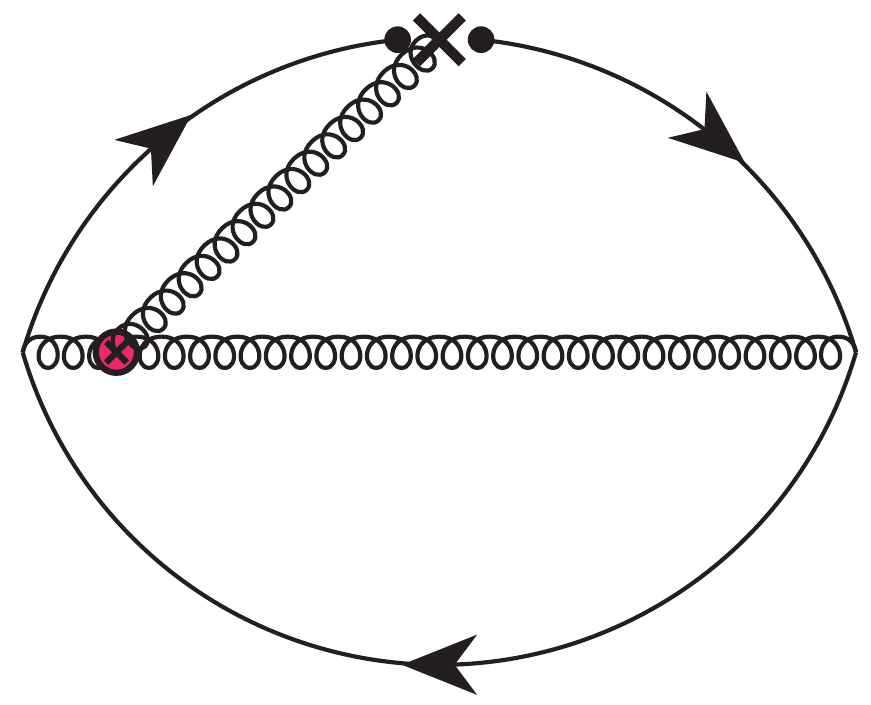}}}~
\\
\subfigure[(e--1)]{
\scalebox{0.12}{\includegraphics{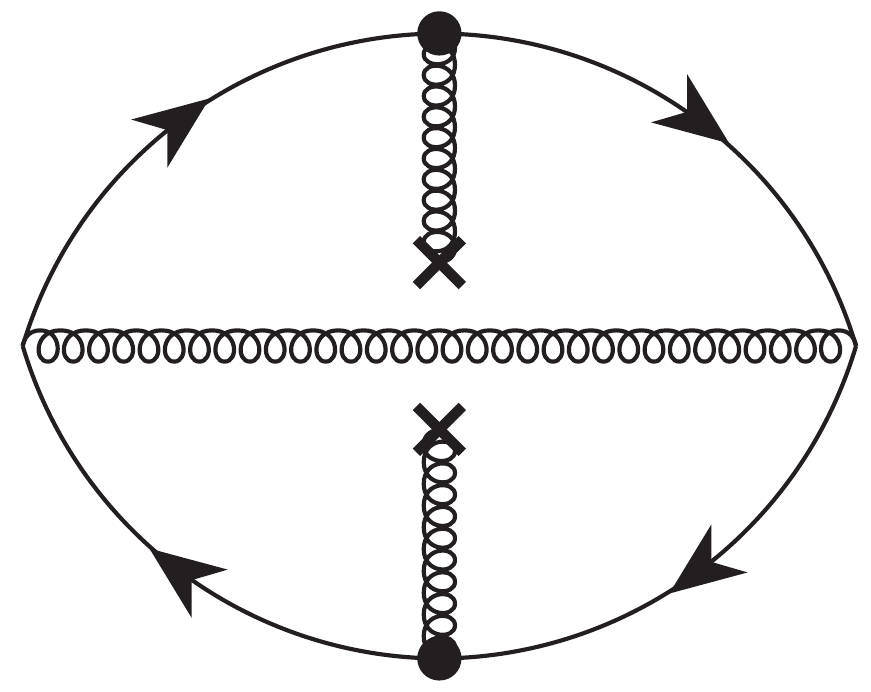}}}~
\subfigure[(e--2)]{
\scalebox{0.12}{\includegraphics{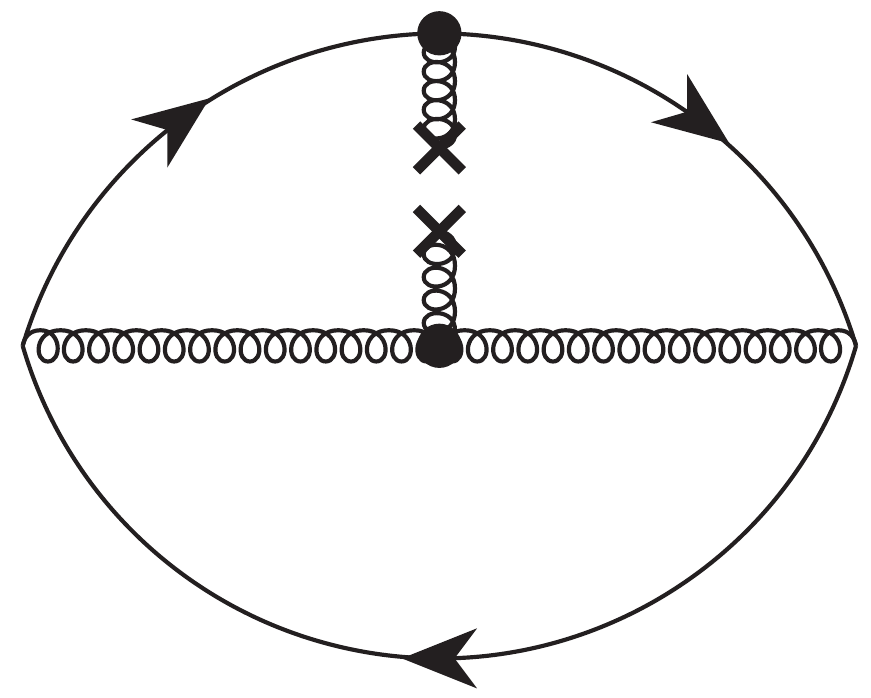}}}
\end{center}
\caption{Feynman diagrams for the single-gluon hybrid state: (a) and (b--i) are proportional to $\alpha_s \times g_s^0$; (c--i) and (d--i) are proportional to $\alpha_s \times g_s^1$; (e--i) are proportional to $\alpha_s \times g_s^2$.}
\label{fig:feynman}
\end{figure}

In the present study we have considered the Feynman diagrams depicted in Fig.~\ref{fig:feynman}, and calculated $\rho_{\rm OPE}(s)$ up to the dimension eight condensates. The gluon field strength tensor $G^n_{\mu\nu}$ is defined as
\begin{equation}
G^n_{\mu\nu} = \partial_\mu A_\nu^n  -  \partial_\nu A_\mu^n  +  g_s f^{npq} A_{p,\mu} A_{q,\nu} \, ,
\end{equation}
which can be separated into the former two terms (represented by the single-gluon-line) and the third term (represented by the double-gluon-line with a red vertex, {\it e.g.}, see the diagram depicted in Fig.~\ref{fig:feynman}(c--3)). In the calculations we have considered the perturbative term, the quark condensates, the quark-gluon mixed condensates, the two-gluon condensate, the three-gluon condensate, and their combinations. We have taken into account all the diagrams proportional to $\alpha_s \times g_s^0$ and $\alpha_s \times g_s^1$, but we have taken into account only three diagrams proportional to $\alpha_s \times g_s^2$. We summarize the obtained OPE spectral densities in Appendix~\ref{app:ope}. Especially, the one extracted from the current $J^{\mu}_{1^{-+}}$ with the quark-gluon content $\bar s s g$ is
\begin{eqnarray}
&& \Pi^{\mu}_{1^{-+}} {\left( {M_B^2,{s_0}} \right)}
\label{eq:ope}
\\ \nonumber
&=& \int_{4m_s^2}^{{s_0}} {(\frac{{{s^3}{\alpha _s}}}{{60{\pi ^3}}}}  - \frac{{m_s^2{s^2}{\alpha _s}}}{{3{\pi ^3}}}+s(\frac{{\left\langle {{\alpha _s}GG} \right\rangle }}{{36{\pi ^2}}}
\\ \nonumber&&
 +\frac{{13\left\langle {{\alpha _s}GG} \right\rangle {\alpha _s}}}{{432{\pi ^3}}} + \frac{{8{m_s}\left\langle {\bar ss} \right\rangle {\alpha _s}}}{{9\pi }})+\frac{{\left\langle {g_s^3{G^3}} \right\rangle }}{{32{\pi ^2}}}
\\ \nonumber&&
 - \frac{{3\left\langle {{\alpha _s}GG} \right\rangle m_s^2{\alpha _s}}}{{64{\pi ^3}}} -\frac{{3{m_s}\left\langle {{g_s}\bar s\sigma Gs} \right\rangle {\alpha _s}}}{{4\pi }}) \times {e^{ - s/M_B^2}}ds
\\ \nonumber
&+&(\frac{{{{\left\langle {{\alpha _s}GG} \right\rangle }^2}}}{{3456{\pi ^2}}} - \frac{{\left\langle {g_s^3{G^3}} \right\rangle m_s^2}}{{16{\pi ^2}}}-\frac{2}{9}\left\langle {{\alpha _s}GG} \right\rangle {m_s}\left\langle {\bar ss} \right\rangle
\\ \nonumber &&
 + \frac{{11\pi \left\langle {\bar ss} \right\rangle \left\langle {{g_s}\bar s\sigma Gs} \right\rangle {\alpha _s}}}{9}) \, .
\end{eqnarray}
The one with the quark-gluon content $\bar q q g$ ($q=u/d$) can be easily derived by replacing $m_s \rightarrow 0$, $\left\langle {\bar ss} \right\rangle \rightarrow \left\langle {\bar qq} \right\rangle$, and $\left\langle {{g_s}\bar s\sigma Gs} \right\rangle \rightarrow \left\langle {{g_s}\bar q\sigma Gq} \right\rangle$. Note that we do not differentiate the up and down quarks in the calculations, so the states in the same isospin multiplet have the same extracted hadron mass, {\it e.g.}, the two $\bar q q g$ states ($q=u/d$) with the quantum numbers $I^GJ^{PC} = 0^+1^{-+}$ and $1^-1^{-+}$, coupled by the same current $J^{\mu}_{1^{-+}}$, have the same extracted hadron mass.

We use the spectral density $\rho^{\bar s s g}_{1^{-+}}(s)$ given in Eq.~(\ref{eq:ope}) to perform numerical analyses. It is extracted from the current $J^{\mu}_{1^{-+}}$ with the quark-gluon content $\bar s s g$, which couples to the state
\begin{equation}
X_{1^{-+}} \equiv |\bar s s g; 1^{-+} \rangle \, .
\end{equation}
We shall use the following values for various QCD parameters at the renormalization scale 2~GeV and the QCD scale $\Lambda_{\rm QCD} = 300$~MeV~\cite{pdg,Ovchinnikov:1988gk,Yang:1993bp,Ellis:1996xc,Ioffe:2002be,Jamin:2002ev,Gimenez:2005nt,Narison:2011xe,Narison:2018dcr}:
\begin{eqnarray}
\nonumber \alpha_s(Q^2) &=& {4\pi \over 11 \ln(Q^2/\Lambda_{\rm QCD}^2)} \, ,
\\ \nonumber \langle\bar qq \rangle &=& -(0.240 \pm 0.010)^3 \mbox{ GeV}^3 \, ,
\\ \nonumber \langle\bar ss \rangle &=& (0.8\pm 0.1)\times \langle\bar qq \rangle \, ,
\\ \langle g_s\bar q\sigma G q\rangle &=& (0.8 \pm 0.2)\times\langle\bar qq\rangle \mbox{ GeV}^2 \, ,
\label{eq:condensate}
\\ \nonumber \langle g_s\bar s\sigma G s\rangle &=&  (0.8 \pm 0.2)\times\langle\bar ss\rangle \, ,
\\ \nonumber \langle \alpha_s GG\rangle &=& (6.35 \pm 0.35) \times 10^{-2} \mbox{ GeV}^4 \, ,
\\ \nonumber \langle g_s^3G^3\rangle &=& (8.2 \pm 1.0) \times \langle \alpha_s GG\rangle  \mbox{ GeV}^2 \, ,
\\ \nonumber m_s &=& 93 ^{+11}_{-~5} \mbox{ MeV} \, .
\end{eqnarray}
Note that the value of the gluon condensate $\langle \alpha_s GG\rangle$ is taken from Ref.~\cite{Narison:2018dcr}, which was written in 2018.

The mass $M_{1^{-+}}$ calculated by Eq.~(\ref{eq:LSR}) depends on two free parameters: the Borel mass $M_B$ and the threshold value $s_0$. We shall determine their proper working regions through three criteria: a) the sufficiently good  convergence of OPE, b) the sufficiently large pole contribution, and c) the sufficiently weak dependence of the mass $M_{1^{-+}}$ on these two parameters.

\begin{figure}[hbtp]
\begin{center}
\includegraphics[width=0.45\textwidth]{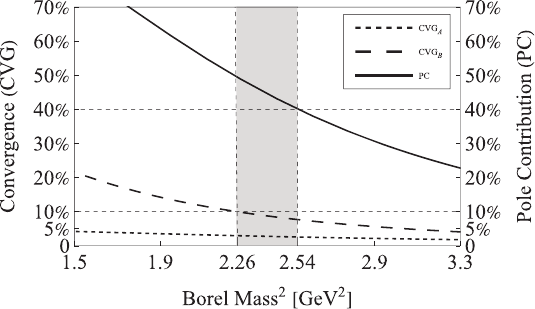}
\caption{CVG$_{A/B}$ and PC with respect to the Borel mass $M_B$, when setting $s_0 = 6.2$~GeV$^2$. These curves are obtained using the spectral density $\rho^{\bar s s g}_{1^{-+}}(s)$ extracted from the current $J^{\mu}_{1^{-+}}$ with the quark-gluon content $\bar s s g$.}
\label{fig:cvgpole}
\end{center}
\end{figure}

In order to have the sufficiently good convergence of OPE, we require the $\alpha_s \times g_s^2$ terms to be less than 5\% and the $D=6+8$ terms to be less than 10\%:
\begin{eqnarray}
\mbox{CVG}_A &\equiv& \left|\frac{ \Pi^{g_s^{n=4}}(\infty, M_B^2) }{ \Pi(\infty, M_B^2) }\right| \leq 5\% \, ,
\\
\mbox{CVG}_B &\equiv& \left|\frac{ \Pi^{{\rm D=6+8}}(\infty, M_B^2) }{ \Pi(\infty, M_B^2) }\right| \leq 10\% \, .
\end{eqnarray}
As depicted in Fig.~\ref{fig:cvgpole}, we determine the minimum Borel mass to be $M_B^2 \geq 2.26$~GeV$^2$, when setting $s_0 = 6.2$~GeV$^2$.

In order to have the sufficiently large pole contribution, we require
\begin{equation}
\mbox{PC} \equiv \left|\frac{ \Pi(s_0, M_B^2) }{ \Pi(\infty, M_B^2) }\right| \geq 40\% \, .
\end{equation}
As depicted in Fig.~\ref{fig:cvgpole}, we determine the maximum Borel mass to be $M_B^2 \leq 2.54$~GeV$^2$, when setting $s_0 = 6.2$~GeV$^2$.

\begin{figure*}[hbtp]
\begin{center}
\includegraphics[width=0.4\textwidth]{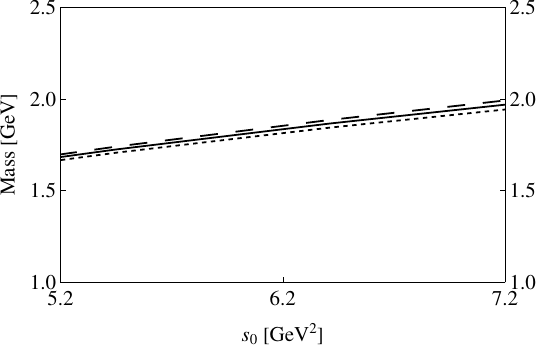}
~~~~~~~~~~
\includegraphics[width=0.4\textwidth]{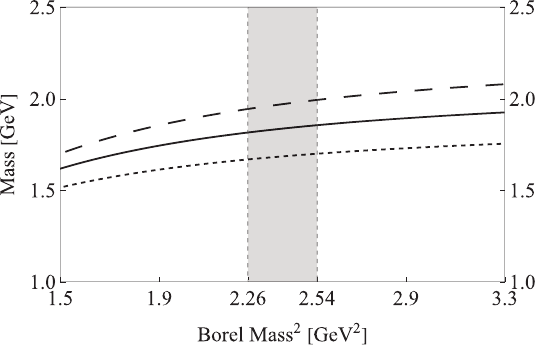}
\caption{Mass of the single-gluon hybrid state $|\bar s s g; 1^{-+} \rangle$ with respect to the threshold value $s_0$ (left) and the Borel mass $M_B$ (right). In the left panel the dotted, solid, and dashed curves are obtained by setting $M_B^2 = 2.26$~GeV$^2$, $2.40$~GeV$^2$, and $2.54$~GeV$^2$, respectively. In the right panel the dotted, solid, and dashed curves are obtained by setting $s_0 = 5.2$~GeV$^2$, $6.2$~GeV$^2$, and $7.2$~GeV$^2$, respectively. These curves are obtained using the spectral density $\rho^{\bar s s g}_{1^{-+}}(s)$ extracted from the current $J^{\mu}_{1^{-+}}$ with the quark-gluon content $\bar s s g$.}
\label{fig:mass}
\end{center}
\end{figure*}

Altogether the Borel window is determined to be $2.26$~GeV$^2 \leq M_B^2 \leq 2.54$~GeV$^2$, when setting $s_0 = 6.2$~GeV$^2$. Note that this Borel window is not so wide, and it was pointed in Ref.~\cite{Matheus:2007ta} that the narrow Borel window may indicate that the understanding of this state as a particle has limitations, so further studies on the hybrid states/particles are crucially demanded. We further change $s_0$ and find that there are non-vanishing Borel windows for $s_0 \geq s^{\rm min}_0 = 5.1$~GeV$^2$. We choose $s_0$ to be about 10\% larger and determine its working region to be $5.2$~GeV$^2 \leq s_0 \leq 7.2$~GeV$^2$, where we calculate the mass and decay constant of the single-gluon hybrid state $X_{1^{-+}} \equiv |\bar s s g; 1^{-+} \rangle$ to be
\begin{eqnarray}
M_{|\bar s s g; 1^{-+} \rangle} &=& 1.84^{+0.14}_{-0.15}{\rm~GeV} \, ,
\label{eq:ssgmass}
\\
f_{|\bar s s g; 1^{-+} \rangle} &=& 0.300^{+0.063}_{-0.058}{\rm~GeV}^4 \, .
\end{eqnarray}
Their uncertainties come from $M_B$ and $s_0$ as well as various QCD parameters listed in Eqs.~(\ref{eq:condensate}). We show the mass $M_{|\bar s s g; 1^{-+} \rangle}$ in Fig.~\ref{fig:mass} with respect to the threshold value $s_0$ and the Borel mass $M_B$. In the left panel, the mass dependence on $s_0$ is moderate inside the working region $5.2$~GeV$^2 \leq s_0 \leq 7.2$~GeV$^2$. In the right panel, the mass curves are sufficiently stable inside the Borel window $2.26$~GeV$^2 \leq M_B^2 \leq 2.54$~GeV$^2$.

Similarly, we perform numerical analyses using the other single-gluon hybrid currents with the quark-gluon contents $\bar q q g$ ($q=u/d$) and $\bar s s g$. The obtained results are summarized in Table~\ref{tab:mass1} and Table~\ref{tab:mass2}. Especially, we use the current $J^{\mu}_{1^{-+}}$ with the quark-gluon content $\bar q q g$ ($q=u/d$) to calculate the mass and decay constant of the single-gluon hybrid state $|\bar q q g;1^{-+}\rangle$ to be
\begin{eqnarray}
M_{|\bar q q g;1^{-+}\rangle} &=& 1.67^{+0.15}_{-0.17}{\rm~GeV} \, ,
\\
f_{|\bar q q g;1^{-+}\rangle} &=& 0.243^{+0.057}_{-0.052}{\rm~GeV}^4 \, .
\label{eq:qqgf}
\end{eqnarray}

\begin{table*}[hbtp]
\begin{center}
\renewcommand{\arraystretch}{1.5}
\caption{QCD sum rule results for the single-gluon hybrid states $|\bar q q g; J^{PC}\rangle$, extracted from the single-gluon hybrid currents given in Eqs.~(\ref{def:B1mm}-\ref{def:At1mm}) and Eqs.~(\ref{def:A0pp}-\ref{def:Bt2pp}) with the quark-gluon contents $\bar q q g$ ($q=u/d$). The results for the isoscalar state $|\bar q q g; 0^GJ^{PC}\rangle$ and the isovector state $|\bar q q g; 1^GJ^{PC}\rangle$ within the same isospin multiplet are the same as each other.}
\begin{tabular}{c|c|c|c|c|c|c|c}
\hline\hline
~\multirow{2}{*}{State [$J^{PC}$]}~ & ~~~~\multirow{2}{*}{Current}~~~~ & ~\multirow{2}{*}{~$s_0^{min}~[{\rm GeV}^2]$~}~ & \multicolumn{2}{c|}{Working Regions} & ~~\multirow{2}{*}{Pole~[\%]}~~ & ~\multirow{2}{*}{~Mass~[GeV]~}~&~~~\multirow{2}{*}{Decay Constant}~~~
\\ \cline{4-5}
&  &  & ~~$M_B^2~[{\rm GeV}^2]$~~ & ~~$s_0~[{\rm GeV}^2]$~~ && &
\\ \hline\hline
$|\bar q q g; 1^{--}\rangle$ & $J^{\alpha\beta}_{1^{--}}$                             &  4.2   &  $2.03$--$2.48$   &  $5.5$   &  $40$--$54$  &  $1.80^{+0.13}_{-0.16}$ &$0.051^{+0.004}_{-0.004}$~GeV$^3$
\\
$|\bar q q g; 1^{+-}\rangle$ & $\tilde{J}^{\alpha\beta}_{1^{+-}}$                     & 16.2   &  $3.61$--$4.58$   &  $18.0$  &  $40$--$53$  &  $4.05^{+0.24}_{-0.12}$ &$0.063^{+0.020}_{-0.020}$~GeV$^3$
\\
$|\bar q q g; 1^{+-}\rangle$ & $ J^{\alpha\beta}_{1^{+-}}$                            &  5.0   &  $2.29$--$2.45$   &  $5.5$   &  $40$--$45$  &  $1.84^{+0.12}_{-0.14}$ &$0.049^{+0.004}_{-0.004}$~GeV$^3$
\\
$|\bar q q g; 1^{--}\rangle$ & $\tilde{J}^{\alpha\beta}_{1^{--}}$                     &  16.3  &  $3.52$--$4.56$   &  $18.0$  &  $40$--$53$  &  $4.09^{+0.29}_{-0.14}$ &$0.064^{+0.021}_{-0.020}$~GeV$^3$
\\
$|\bar q q g; 0^{++}\rangle$ & $J^{\mu\rightarrow0}_{1^{-+}}$                         &  20.6  &  $5.11$--$6.59$   &  $24.0$  &  $40$--$56$  &  $4.45^{+0.22}_{-0.17}$ &$0.124^{+0.032}_{-0.036}$~GeV$^3$
\\
$|\bar q q g; 0^{-+}\rangle$ & $\tilde{J}^{\mu\rightarrow0}_{1^{++}}$                 & 7.7    &  $3.58$--$3.81$   &  $8.5$   &  $40$--$45$  &  $2.14^{+0.17}_{-0.19}$ &$0.105^{+0.005}_{-0.004}$~GeV$^3$
\\
$|\bar q q g; 0^{--}\rangle$ & $J^{\mu\rightarrow0}_{1^{+-}}$                         &  21.6  &  $5.48$--$6.52$   &  $24.0$  &  $40$--$50$  &  $4.49^{+0.21}_{-0.14}$ &$0.123^{+0.032}_{-0.037}$~GeV$^3$
\\
$|\bar q q g; 0^{+-}\rangle$ & $\tilde{J}^{\mu\rightarrow0}_{1^{--}}$                 & 7.1    &  $3.32$--$3.73$   &  $8.5$   &  $40$--$49$  &  $2.16^{+0.16}_{-0.19}$ &$0.100^{+0.005}_{-0.005}$~GeV$^3$
\\
$|\bar q q g; 1^{-+}\rangle$ & $J^{\mu}_{1^{-+}}$                                     &  4.8   &  $2.19$--$2.28$   &  $5.2$   &  $40$--$43$  &  $1.67^{+0.15}_{-0.17}$ &$0.243^{+0.057}_{-0.052}$~GeV$^4$
\\
$|\bar q q g; 1^{++}\rangle$ & $\tilde{J}^{\mu}_{1^{++}}$                             &  13.8  &  $3.59$--$4.10$   &  $15.0$  &  $40$--$48$  &  $3.54^{+0.16}_{-0.12}$ &$1.370^{+0.494}_{-0.450}$~GeV$^4$
\\
$|\bar q q g; 1^{+-}\rangle$ & $J^{\mu}_{1^{+-}}$                                     & 4.6    &  $2.10$--$2.27$   &  $5.2$   &  $40$--$46$  &  $1.68^{+0.14}_{-0.16}$ &$0.242^{+0.055}_{-0.051}$~GeV$^4$
\\
$|\bar q q g; 1^{--}\rangle$ & $\tilde{J}^{\mu}_{1^{--}}$                             & 13.7   &  $3.57$--$4.10$   &  $15.0$  &  $40$--$49$  &  $3.53^{+0.16}_{-0.12}$ &$1.366^{+0.493}_{-0.450}$~GeV$^4$
\\
$|\bar q q g; 0^{++}\rangle$ & $J_{0^{++}}$                                           &  11.1  &  $3.48$--$3.91$   &  $12.5$  &  $40$--$49$  &  $2.94^{+0.20}_{-0.25}$ &$2.893^{+1.029}_{-0.948}$~GeV$^4$
\\
$|\bar q q g; 0^{-+}\rangle$ & $J_{0^{-+}}$                                           &  11.1  &  $3.47$--$3.92$   &  $12.5$  &  $40$--$49$  &  $2.93^{+0.20}_{-0.25}$ &$2.882^{+1.026}_{-0.945}$~GeV$^4$
\\
$|\bar q q g; 1^{++}\rangle$ & $J^{\alpha\beta}_{1^{++}}$                             &  5.8   &  $1.84$--$2.06$   &  $6.5$   &  $40$--$48$  &  $2.11^{+0.17}_{-0.21}$ &$0.056^{+0.012}_{-0.013}$~GeV$^3$
\\
$|\bar q q g; 1^{-+}\rangle$ & $\tilde{J}^{\alpha\beta}_{1^{-+}}$                     &  5.5   &  $1.81$--$2.00$   &  $6.2$   &  $40$--$48$  &  $2.00^{+0.13}_{-0.16}$ &$0.055^{+0.007}_{-0.008}$~GeV$^3$
\\
$|\bar q q g; 1^{-+}\rangle$ & $J^{\alpha\beta}_{1^{-+}}$                             &  5.5   &  $1.81$--$2.00$   &  $6.2$   &  $40$--$48$  &  $2.00^{+0.13}_{-0.16}$ &$0.055^{+0.007}_{-0.008}$~GeV$^3$
\\
$|\bar q q g; 1^{++}\rangle$ & $\tilde{J}^{\alpha\beta}_{1^{++}}$                     &  5.8   &  $1.84$--$2.06$   &  $6.5$   &  $40$--$48$  &  $2.11^{+0.17}_{-0.21}$ &$0.056^{+0.012}_{-0.013}$~GeV$^3$
\\
$|\bar q q g; 2^{++}\rangle$ & $J^{\alpha_1\beta_1,\alpha_2\beta_2}_{2^{++}}$         &  8.6   &  $3.11$--$3.37$   &  $9.5$   &  $40$--$46$  &  $2.44^{+0.20}_{-0.24}$ & --
\\
$|\bar q q g; 2^{-+}\rangle$ & $\tilde{J}^{\alpha_1\beta_1,\alpha_2\beta_2}_{2^{-+}}$ &  12.7  &  $2.54$--$3.60$   &  $14.0$  &  $40$--$54$  &  $3.68^{+0.62}_{-0.18}$ & --
\\
$|\bar q q g; 2^{-+}\rangle$ & $J^{\alpha_1\beta_1,\alpha_2\beta_2}_{2^{-+}}$         &  8.3   &  $3.07$--$3.41$   & $9.5$    &  $40$--$48$  &  $2.40^{+0.21}_{-0.25}$ & --
\\
$|\bar q q g; 2^{++}\rangle$ & $\tilde{J}^{\alpha_1\beta_1,\alpha_2\beta_2}_{2^{++}}$ &  11.7  &  $2.47$--$3.70$   &  $14.0$  &  $40$--$63$  &  $3.46^{+0.27}_{-0.11}$ & --
\\ \hline\hline
\end{tabular}
\label{tab:mass1}
\end{center}
\end{table*}

\begin{table*}[hbtp]
\begin{center}
\renewcommand{\arraystretch}{1.5}
\caption{QCD sum rule results for the single-gluon hybrid states $|\bar s s g; J^{PC}\rangle$, extracted from the single-gluon hybrid currents given in Eqs.~(\ref{def:B1mm}-\ref{def:At1mm}) and Eqs.~(\ref{def:A0pp}-\ref{def:Bt2pp}) with the quark-gluon contents $\bar s s g$.}
\begin{tabular}{c|c|c|c|c|c|c|c}
\hline\hline
~\multirow{2}{*}{State [$J^{PC}$]}~ & ~~~~\multirow{2}{*}{Current}~~~~ & ~\multirow{2}{*}{~$s_0^{min}~[{\rm GeV}^2]$~}~ & \multicolumn{2}{c|}{Working Regions} & ~~\multirow{2}{*}{Pole~[\%]}~~ & ~\multirow{2}{*}{~Mass~[GeV]~}~&~~~\multirow{2}{*}{Decay Constant}~~~
\\ \cline{4-5}
&  &  & ~~$M_B^2~[{\rm GeV}^2]$~~ & ~~$s_0~[{\rm GeV}^2]$~~ && &
\\ \hline\hline
$|\bar s s g; 1^{--}\rangle$ & $J^{\alpha\beta}_{1^{--}}$                                   &  4.3   &  $2.07$--$2.80$   &  $6.5$   &  $40$--$63$  &  $1.94^{+0.20}_{-0.21}$ &$0.054^{+0.013}_{-0.016}$~GeV$^3$
\\
$|\bar s s g; 1^{+-}\rangle$ & $\tilde{J}^{\alpha\beta}_{1^{+-}}$                           & 16.2   &  $3.60$--$5.40$   &  $20.0$  &  $40$--$65$  &  $4.06^{+0.26}_{-0.16}$ &$0.071^{+0.019}_{-0.020}$~GeV$^3$
\\
$|\bar s s g; 1^{+-}\rangle$ & $ J^{\alpha\beta}_{1^{+-}}$                                  &  5.9   &  $2.54$--$2.72$   &  $6.5$   &  $40$--$45$  &  $2.01^{+0.17}_{-0.20}$ &$0.050^{+0.005}_{-0.006}$~GeV$^3$
\\
$|\bar s s g; 1^{--}\rangle$ & $\tilde{J}^{\alpha\beta}_{1^{--}}$                           &  16.9  &  $3.73$--$5.30$   &  $20.0$  &  $40$--$61$  &  $4.12^{+0.26}_{-0.13}$ &$0.070^{+0.019}_{-0.020}$~GeV$^3$
\\
$|\bar s s g; 0^{++}\rangle$ & $J^{\mu\rightarrow0}_{1^{-+}}$                               &  20.7  &  $5.18$--$7.35$   &  $26.0$  &  $40$--$63$  &  $4.50^{+0.23}_{-0.22}$ &$0.136^{+0.030}_{-0.034}$~GeV$^3$
\\
$|\bar s s g; 0^{-+}\rangle$ & $\tilde{J}^{\mu\rightarrow0}_{1^{++}}$                       &  7.2   &  $3.45$--$4.08$   &  $9.5$   &  $40$--$53$  &  $2.26^{+0.21}_{-0.24}$ &$0.107^{+0.007}_{-0.005}$~GeV$^3$
\\
$|\bar s s g; 0^{--}\rangle$ & $J^{\mu\rightarrow0}_{1^{+-}}$                               &  21.6  &  $5.36$--$7.23$   &  $26.0$  &  $40$--$59$  &  $4.57^{+0.22}_{-0.19}$ &$0.134^{+0.031}_{-0.035}$~GeV$^3$
\\
$|\bar s s g; 0^{+-}\rangle$ & $\tilde{J}^{\mu\rightarrow0}_{1^{--}}$                       & 7.5    &  $3.41$--$3.98$   &  $9.5$   &  $40$--$52$  &  $2.30^{+0.20}_{-0.24}$ &$0.101^{+0.007}_{-0.006}$~GeV$^3$
\\
$|\bar s s g; 1^{-+}\rangle$ & $J^{\mu}_{1^{-+}}$                                           &  5.1   &  $2.26$--$2.54$   &  $6.2$   &  $40$--$49$  &  $1.84^{+0.14}_{-0.15}$ &$0.300^{+0.063}_{-0.058}$~GeV$^4$
\\
$|\bar s s g; 1^{++}\rangle$ & $\tilde{J}^{\mu}_{1^{++}}$                                   &  14.1  &  $3.64$--$4.80$   &  $17.0$  &  $40$--$58$  &  $3.65^{+0.17}_{-0.17}$ & $1.678^{+0.530}_{-0.502}$~GeV$^4$
\\
$|\bar s s g; 1^{+-}\rangle$ & $J^{\mu}_{1^{+-}}$                                           & 3.9    &  $1.85$--$2.43$   &  $6.0$   &  $40$--$62$  &  $1.82^{+0.13}_{-0.15}$ &$0.278^{+0.059}_{-0.056}$~GeV$^4$
\\
$|\bar s s g; 1^{--}\rangle$ & $\tilde{J}^{\mu}_{1^{--}}$                                   & 13.8   &  $3.50$--$4.80$   &  $17.0$  &  $40$--$61$  &  $3.64^{+0.17}_{-0.17}$ &$1.662^{+0.526}_{-0.498}$~GeV$^4$
\\
$|\bar s s g; 0^{++}\rangle$ & $J_{0^{++}}$                                                 &  11.5  &  $3.53$--$4.33$   &  $14.0$  &  $40$--$55$  &  $3.11^{+0.22}_{-0.27}$ &$3.535^{+1.338}_{-1.242}$~GeV$^4$
\\
$|\bar s s g; 0^{-+}\rangle$ & $J_{0^{-+}}$                                                 &  11.3  &  $3.51$--$4.36$   &  $14.0$  &  $40$--$56$  &  $3.08^{+0.23}_{-0.28}$ &$3.509^{+1.328}_{-1.233}$~GeV$^4$
\\
$|\bar s s g; 1^{++}\rangle$ & $J^{\alpha\beta}_{1^{++}}$                                   &  6.6   &  $1.95$--$2.27$   &  $7.5$   &  $40$--$51$  &  $2.34^{+0.14}_{-0.16}$ &$0.061^{+0.012}_{-0.014}$~GeV$^3$
\\
$|\bar s s g; 1^{-+}\rangle$ & $\tilde{J}^{\alpha\beta}_{1^{-+}}$                           &  5.5   &  $1.82$--$2.25$   &  $7.0$   &  $40$--$57$  &  $2.08^{+0.18}_{-0.24}$ &$0.061^{+0.010}_{-0.010}$~GeV$^3$
\\
$|\bar s s g; 1^{-+}\rangle$ & $J^{\alpha\beta}_{1^{-+}}$                                   &  5.5   &  $1.82$--$2.25$   &  $7.0$   &  $40$--$57$  &  $2.08^{+0.18}_{-0.24}$ &$0.061^{+0.010}_{-0.010}$~GeV$^3$
\\
$|\bar s s g; 1^{++}\rangle$ & $\tilde{J}^{\alpha\beta}_{1^{++}}$                           &  6.6   &  $1.95$--$2.27$   &  $7.5$   &  $40$--$51$  &  $2.34^{+0.14}_{-0.16}$ &$0.061^{+0.012}_{-0.014}$~GeV$^3$
\\
$|\bar s s g; 2^{++}\rangle$ & $J^{\alpha_1\beta_1,\alpha_2\beta_2}_{2^{++}}$               &  9.2   &  $3.22$--$3.60$   &  $10.5$  &  $40$--$49$  &  $2.59^{+0.19}_{-0.23}$ & --
\\
$|\bar s s g; 2^{-+}\rangle$ & $\tilde{J}^{\alpha_1\beta_1,\alpha_2\beta_2}_{2^{-+}}$       &  13.4  &  $2.55$--$4.29$   &  $16.0$  &  $40$--$66$  &  $3.72^{+0.72}_{-0.13}$ & --
\\
$|\bar s s g; 2^{-+}\rangle$ & $J^{\alpha_1\beta_1,\alpha_2\beta_2}_{2^{-+}}$               &  8.1   &  $3.04$--$3.72$   & $10.5$   &  $40$--$56$  &  $2.51^{+0.20}_{-0.24}$ & --
\\
$|\bar s s g; 2^{++}\rangle$ & $\tilde{J}^{\alpha_1\beta_1,\alpha_2\beta_2}_{2^{++}}$       &  11.8  &  $2.36$--$4.47$   &  $16.0$  &  $40$--$78$  &  $3.54^{+0.42}_{-0.16}$ & --
\\ \hline\hline
\end{tabular}
\label{tab:mass2}
\end{center}
\end{table*}

\section{Decay properties of the $J^{PC} = 1^{-+}$ hybrid states}
\label{sec:decay}

\begin{figure*}[hbtp]
\begin{center}
\subfigure[(a)]{\includegraphics[width=0.3\textwidth]{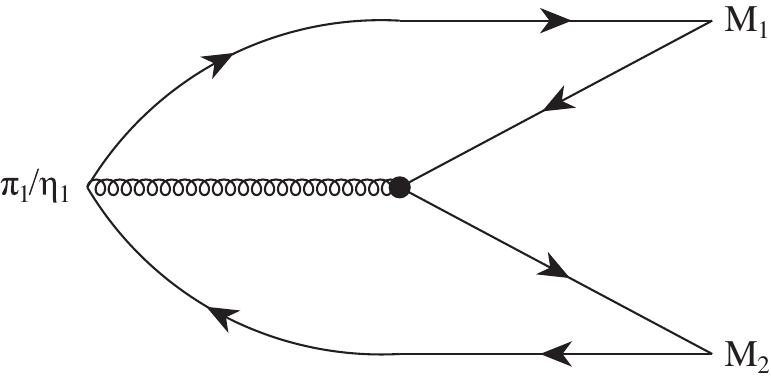}}
~~~~~
\subfigure[(b)]{\includegraphics[width=0.3\textwidth]{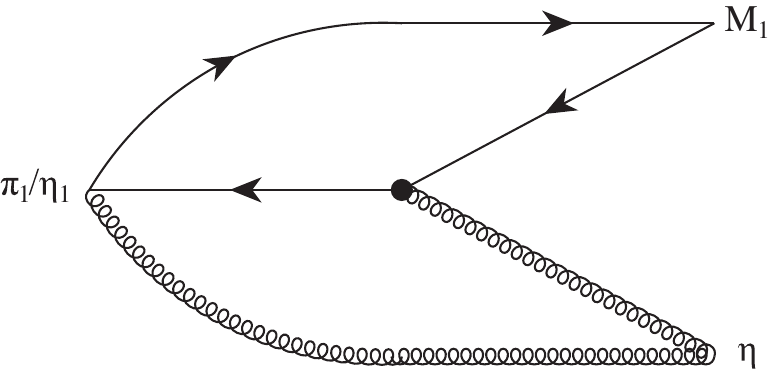}}
~~~~~
\subfigure[(c)]{\includegraphics[width=0.3\textwidth]{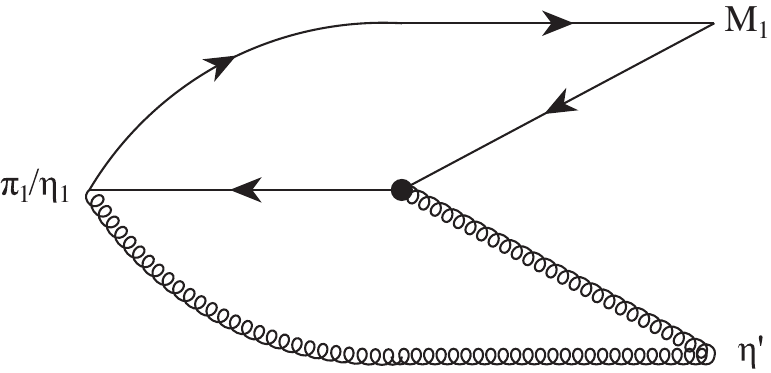}}
\end{center}
\caption{Decay mechanisms of the single-gluon hybrid states through (a) the normal process with one quark-antiquark pair excited from the valence gluon, and (b,c) the abnormal processes with the $\eta/\eta^{\prime}$ produced by the QCD axial anomaly.}
\label{fig:decay}
\end{figure*}

In this section we systematically study the decay properties of the $J^{PC} = 1^{-+}$ hybrid states, whose masses and decay constants have been calculated in the previous section. Since we do not differentiate the up and down quarks within the QCD sum rule method, the masses and decay constants of the states in the same isospin multiplet are calculated to be the same:
\begin{eqnarray}
\nonumber M_{|\bar q q g;1^-1^{-+}\rangle} = M_{|\bar q q g;0^+1^{-+}\rangle} &=&1.67^{+0.15}_{-0.17}{\rm~GeV} \, ,
\\ \nonumber f_{|\bar q q g;1^-1^{-+}\rangle} = f_{|\bar q q g;0^+1^{-+}\rangle} &=& 0.243^{+0.057}_{-0.052}{\rm~GeV}^4 \, ,
\\ \nonumber M_{|\bar s s g;0^+1^{-+}\rangle} &=& 1.84^{+0.14}_{-0.15}{\rm~GeV} \, ,
\\ \nonumber f_{|\bar s s g;0^+1^{-+}\rangle} &=& 0.300^{+0.063}_{-0.058}{\rm~GeV}^4 \, .
\end{eqnarray}
As shown in Fig.~\ref{fig:decay}(a), a single-gluon hybrid state can decay after exciting one $\bar q q/\bar s s$ pair from the valence gluon, followed by reorganizing two color-octet $\bar q q/\bar s s$ pairs into two color-singlet mesons. This decay process has been systematically studied in Refs.~\cite{Chen:2010ic,Huang:2010dc} for the $J^{PC} = 1^{-+}$ hybrid states through the QCD sum rule method, and in the present study we update these calculations. Besides the ``normal'' decay process depicted in Fig.~\ref{fig:decay}(a), there also exist the ``abnormal'' decay processes depicted in Figs.~\ref{fig:decay}(b,c), where the $\eta^{(\prime)}$ mesons are produced by the QCD axial anomaly. These abnormal decay processes have been systematically studied in Ref.~\cite{Chen:2022qpd}, and in the present study we also update these calculations.

We shall use the decay modes $\pi_1 \equiv |\bar q q g;1^-1^{-+}\rangle \rightarrow \rho \pi$ and $\eta_1 \equiv |\bar s s g;0^+1^{-+}\rangle \xrightarrow{c} \eta \eta^\prime$ as two examples to separately study the normal and abnormal decay processes in the following subsections.

\subsection{Normal decay process}

In this subsection we use the decay mode
\begin{equation}
\pi_1 \equiv |\bar q q g;1^-1^{-+}\rangle \rightarrow \rho \pi \, ,
\end{equation}
as an example to study the normal decay process depicted in Fig.~\ref{fig:decay}(a) through the three-point correlation function
\begin{eqnarray}
T_{\mu\nu}(p, k, q) &=& \int d^4x d^4y e^{i k x} e ^{i q y} \times
\\ \nonumber
&& ~~~~~~ \langle0|{\mathbb T}[ J_\nu^{\rho^-}(x) J_5^{\pi^+}(y) J^{\mu \dagger}_{1^{-+}}(0) ] |0\rangle \, ,
\end{eqnarray}
where $p$, $k$ and $q$ are the momenta of $\pi_1 \equiv |\bar q q g;1^-1^{-+}\rangle$, $\rho^-$, and $\pi^+$, respectively. The current $J^{\mu}_{1^{-+}}$ has been defined in Eq.~(\ref{def:A1mp}), and we select the isovector neutral-charged one
\begin{equation}
J^{\mu}_{1^{-+}} \rightarrow {1\over\sqrt2}\left( \bar u_a \lambda_n^{ab} \gamma_\beta u_b - \bar d_a \lambda_n^{ab} \gamma_\beta d_b \right)g_s G_n^{\mu\beta} \, .
\end{equation}
The negative-charged vector current $J_\mu^{\rho^-} = \bar u \gamma_\mu d$ and the positive-charged pseudoscalar current $J_5^{\pi^+} = \bar d \gamma_5 u$ respectively couple to the vector meson $\rho^-$ and the pseudoscalar meson $\pi^+$ through
\begin{eqnarray}
\langle0| J_\mu^{\rho^-} |\rho^-(k,\epsilon)\rangle &=& m_\rho f_\rho \epsilon_\mu \, ,
\\
\langle0| J_5^{\pi^+} |\pi(q)\rangle &=& f_\pi^\prime = {2i \langle \bar q q \rangle \over f_\pi} \, ,
\end{eqnarray}
with~\cite{Zhu:1999wg,Reinders:1984sr,Zhu:1998bm}:
\begin{eqnarray}
&& m_\pi = {140 \rm MeV} \, , ~~~ f_\pi = 131 {\rm MeV} \, ,
\\ \nonumber &&
m_\rho = 770 {\rm MeV} \, , ~~~ f_\rho = 220 {\rm MeV} \, .
\end{eqnarray}

At the phenomenological side we write $T_{\mu\nu}(p, k, q)$ as
\begin{eqnarray}
\label{eq:rhopiPH}
T_{\mu\nu}(p, k, q) &=& g_{\rho \pi} \epsilon_{\mu\nu\alpha\beta} {q^\alpha k^\beta}
\\ \nonumber &\times& { f_{\pi_1} f_\rho m_\rho f_\pi^\prime \over (m_{\pi_1}^2 - p^2) (m_\rho^2 - k^2)(m_\pi^2 - q^2)} \, ,
\end{eqnarray}
where the coupling constant $g_{\rho \pi}$ is defined through the Lagrangian
\begin{eqnarray}
{\cal L}=g_{\rho \pi}\epsilon_{\mu\nu\alpha\beta} ~ \pi_1^{0\mu} ~ \partial^\alpha\pi^+ ~ \partial^\beta \rho^{-\nu} + \cdots \, .
\end{eqnarray}

At the QCD side we calculate $T_{\mu\nu}(p, k, q)$ using the method of operator product expansion. We work at the pion pole and choose the terms divergent at the $q^2 \rightarrow 0$ limit to derive
\begin{eqnarray}\label{eq:rhopiOPE}
&& T_{\mu\nu}(p, k, q) = {\epsilon_{\mu\nu\alpha\beta} {q^\alpha k^\beta} \over q^2 } \times
\\ \nonumber && ~~~ \Big ( {\langle g_s \bar q \sigma G q \rangle \over 6\sqrt2 } \big( {3 \over p^2} + { 1 \over k^2 } \big )
- {\langle \bar q q \rangle \langle g_s^2 GG \rangle \over 18 \sqrt2 } \big(  { 1 \over p^4} + { 1 \over k^4} \big ) \Big ) \, .
\end{eqnarray}

We compare Eq.~(\ref{eq:rhopiPH}) and Eq.~(\ref{eq:rhopiOPE}) to calculate the coupling constant $g_{\rho \pi}$. After setting $p^2 = k^2$ and performing the Borel transformation once $\mathcal{B}(p^2 = k^2 \rightarrow T^2)$, we arrive at
\begin{eqnarray}\label{eq:rhopi}
\nonumber && -g_{\rho \pi} { f_{\pi_1} f_\rho m_\rho f_\pi^\prime \over m_\rho^2 - m_{\pi_1}^2 } \Big ( e^{-m_{\pi_1}^2/T^2} - e^{-m_\rho^2/T^2} \Big )
\\ &=& - {2\langle g \bar q \sigma G q \rangle \over 3\sqrt2 } - {\langle \bar q q \rangle \langle g_s^2 GG \rangle \over 9 \sqrt2 } {1\over T^2} \, .
\end{eqnarray}
The formula of the decay width reads
\begin{equation}
\Gamma(\pi_1^0 \rightarrow \rho^+ \pi^- + \rho^- \pi^+) = 2 \times {g_{\rho \pi}^2 \over 12 \pi } |\vec q_\pi|^3 \, ,
\end{equation}
where $\vec q_\pi$ is the three-momentum of the final state $\pi$. Numerically, we obtain
\begin{eqnarray}
g_{\rho \pi} &=&  {4.08^{+2.40}_{-1.83}~\rm GeV^{-1}} \, ,
\\
\Gamma(\pi_1 \rightarrow \rho \pi) &=& {242^{+310}_{-179}~\rm MeV} \, .
\end{eqnarray}

\subsection{Abnormal decay process}

In this subsection we use the decay mode
\begin{equation}
\eta_1 \equiv |\bar s s g;0^+1^{-+}\rangle \xrightarrow{c} \eta \eta^\prime \, ,
\end{equation}
as an example to study the abnormal decay process depicted in Fig.~\ref{fig:decay}(c), with $\eta^\prime$ produced by the QCD $U(1)_A$ anomaly. Before doing this, let us shortly introduce the two-angle mixing formalism to describe the $\eta$ and $\eta^\prime$ mesons~\cite{Leutwyler:1997yr,Kaiser:1998ds,Escribano:2005qq,Escribano:2015nra,Escribano:2015yup,Schechter:1992iz,Kiselev:1992ms,Herrera-Siklody:1997pgy,Bass:2018xmz,Bali:2021qem}:
\begin{eqnarray}
|\eta\rangle &=& \cos\theta_8 |\eta_8\rangle - \sin \theta_0 | \eta_0 \rangle + \cdots \, ,
\\ \nonumber
|\eta^\prime\rangle &=& \sin\theta_8 |\eta_8\rangle + \cos \theta_0 | \eta_0 \rangle + \cdots \, ,
\end{eqnarray}
with
\begin{eqnarray}
|\eta_8 \rangle &=& | u \bar u + d \bar d - 2 s \bar s \rangle/\sqrt6 \, ,
\\ \nonumber
|\eta_0 \rangle &=& | u \bar u + d \bar d + s \bar s \rangle/\sqrt3 \, ,
\end{eqnarray}
and $\cdots$ are contributions from the pseudoscalar glueballs and charmonium, etc.

The octet and singlet axial-vector currents are defined as
\begin{eqnarray}
A_\mu^{8} &=& \left( { \bar u \gamma_\mu \gamma_5 u +  \bar d \gamma_\mu \gamma_5 d - 2 \bar s \gamma_\mu \gamma_5 s }\right)/\sqrt{12} \, ,
\\ \nonumber
A_\mu^{0} &=& \left( { \bar u \gamma_\mu \gamma_5 u +  \bar d \gamma_\mu \gamma_5 d + \bar s \gamma_\mu \gamma_5 s }\right)/\sqrt6 \, .
\end{eqnarray}
These two currents couple to the $\eta$ and $\eta^\prime$ mesons through
\begin{equation}
\langle0| A_\mu^a | P(k) \rangle = i k_\mu f_P^a \, ,
\end{equation}
where $f_P^a$ ($a=8,0;\,P=\eta,\eta^\prime$) is the matrix for the decay constants, defined as
\begin{equation}
\left(\begin{array}{cc}
f_\eta^8 & f_\eta^0
\\
f_{\eta^\prime}^8 & f_{\eta^\prime}^0
\end{array}\right)
=
\left(\begin{array}{cc}
f_8 \cos\theta_8 & - f_0 \sin\theta_0
\\
f_8 \sin\theta_8 &   f_0 \cos\theta_0
\end{array}\right) \, .
\end{equation}

To simply our notations, we further construct the axial-vector currents
\begin{eqnarray}
J_\mu^\eta = A_\mu^8 + t_\eta A_\mu^0 \, ,
\label{def:eta}
\\
J_\mu^{\eta\prime} = A_\mu^8 +  t_{\eta^\prime} A_\mu^0 \, ,
\end{eqnarray}
which couple to the $\eta$ and $\eta^\prime$ mesons through
\begin{eqnarray}
\nonumber \langle0| J_\mu^\eta |\eta(k) \rangle &=& i k_\mu g_\eta \, ,
\\
\langle0| J_\mu^{\eta\prime} |\eta^\prime(k) \rangle &=& i k_\mu g_{\eta^\prime} \, ,
\\ \nonumber
\langle0| J_\mu^\eta |\eta^\prime(k) \rangle &=& \langle0| J_\mu^{\eta\prime} |\eta(k) \rangle = 0 \, ,
\end{eqnarray}
with
\begin{eqnarray}
\nonumber g_\eta &=& f_\eta^8 - f_\eta^0 f_{\eta^\prime}^8/f_{\eta^\prime}^0 \, ,
\\ g_{\eta^\prime} &=& f_{\eta^\prime}^8 - f_{\eta^\prime}^0 f_{\eta}^8/f_{\eta}^0 \, ,
\\ \nonumber t_\eta &=& - f_{\eta^\prime}^8/f_{\eta^\prime}^0 \, ,
\\ \nonumber t_{\eta^\prime} &=& - f_{\eta}^8/f_{\eta}^0 \, .
\end{eqnarray}
We shall use the following values in the calculations~\cite{Ali:1997ex,Feldmann:1997vc}:
\begin{eqnarray}
\nonumber \theta_8 &=& - 22.2^\circ \, ,
\\ \nonumber \theta_0 &=& - 9.1^\circ \, ,
\\ f_8 &=& 168 {\rm~MeV} \, ,
\label{parameter1}
\\ \nonumber f_0 &=& 157 {\rm~MeV} \, .
\end{eqnarray}

The conversion of the gluons into the $\eta$ and $\eta^\prime$ mesons can be described through the QCD $U(1)_A$ anomaly as~\cite{Voloshin:1980zf,Akhoury:1987ed,Castoldi:1988dm,Chao:1989yp,Ball:1995zv,Ali:1997ex}:
\begin{eqnarray}
\langle 0| {\alpha_s \over 4 \pi} G^{\alpha\beta}_n \tilde G_{n,\alpha\beta} | \eta \rangle &=& m^2_{\eta} f_{\eta} \, ,
\label{eq:anomaly1}
\\
\langle 0| {\alpha_s \over 4 \pi} G^{\alpha\beta}_n \tilde G_{n,\alpha\beta} | \eta^\prime \rangle &=& m^2_{\eta^\prime} f_{\eta^\prime} \, ,
\label{eq:anomaly2}
\end{eqnarray}
with
\begin{eqnarray}
f_{\eta} &=& {f_8 \over \sqrt6} \cos\theta_8 - {f_0 \over \sqrt3} \sin\theta_0 \, ,
\\
f_{\eta^\prime} &=& {f_8 \over \sqrt6} \sin\theta_8 + {f_0 \over \sqrt3} \cos\theta_0 \, .
\end{eqnarray}

To study the abnormal decay process $\eta_1 \equiv |\bar s s g;0^+1^{-+}\rangle \xrightarrow{c} \eta \eta^\prime$ depicted in Fig.~\ref{fig:decay}(c), with $\eta^\prime$ produced by the QCD $U(1)_A$ anomaly, we consider the three-point correlation function:
\begin{equation}
T^\prime_{\mu\nu}(p, k, q) = \int d^4x e^{-ikx} \langle 0 | {\mathbb T}[ J^{\mu}_{1^{-+}}(0) J_\nu^{\eta\dagger}(x) ] | \eta^\prime \rangle \, ,
\label{eq:etapi}
\end{equation}
where $p$, $k$, and $q$ are the momenta of $\eta_1 \equiv |\bar s s g;0^+1^{-+}\rangle$, $\eta$, and $\eta^\prime$, respectively. The current $J_\nu^{\eta}$ has been defined in Eq.~(\ref{def:eta}). The current $J^{\mu}_{1^{-+}}$ has been defined in Eq.~(\ref{def:A1mp}), and we set its quark content to be
\begin{equation}
J^{\mu}_{1^{-+}} \rightarrow \bar s_a \lambda_n^{ab} \gamma_\beta s_b g_s G_n^{\mu\beta} \, .
\end{equation}

At the phenomenological side we write $T^\prime_{\mu\nu}(p, k, q)$ as
\begin{equation}
T^\prime_{\mu\nu}(p, k, q) = g_{\eta\eta^\prime} k_\mu k_\nu ~ { f_{\eta_1} g_{\eta} \over (m_{\eta_1}^2 - p^2) (m_{\eta}^2 - k^2)} \, ,
\label{eq:etapi1}
\end{equation}
where the coupling constant $g_{\eta\pi}$ is defined through the Lagrangian
\begin{equation}
\mathcal{L}^\prime = g_{\eta\eta^\prime}~\eta_1^{\mu}~(\partial_\mu \eta)~\eta^\prime \, .
\end{equation}

At the QCD side we calculate $T^\prime_{\mu\nu}(p, k, q)$ using the method of operator product expansion to be
\begin{eqnarray}
&& T^\prime_{\mu\nu}(p, k, q)
\\ \nonumber &=&
\theta_s k_\mu k_\nu ~ \Big( - {2 m^2_{\eta^\prime} f_{\eta^\prime} \over 3 k^2 } - {4\pi^2 m^2_{\eta^\prime} f_{\eta^\prime} m_s \langle \bar s s\rangle \over 3k^6} \Big) \, ,
\end{eqnarray}
where $\theta_s = -{1 / \sqrt3} + { t_\eta / \sqrt6}$ describes the $s \bar s$ component contained in the current $J_\nu^{\eta}$.

After setting $p^2 = k^2$ and performing the Borel transformation once $\mathcal{B}(p^2 = k^2 \rightarrow T^2)$, we arrive at
\begin{eqnarray}
&& \nonumber g_{\eta\eta^\prime} {f_{\eta_1} g_\eta \over m_{\eta_1}^2 - m_\eta^2} \left( e^{-m_{\eta}^2/M_B^2} -  e^{-m_{\eta_1}^2/M_B^2} \right)
\\ &=& {2 \theta_s m^2_{\eta^\prime} f_{\eta^\prime} \over 3 } + {2 \pi^2 \theta_s m^2_{\eta^\prime} f_{\eta^\prime} m_s \langle \bar s s\rangle \over 3} {1 \over M_B^4} \, .
\label{eq:etaetap}
\end{eqnarray}
The formula of the decay width reads
\begin{equation}
\Gamma(\eta_1 \rightarrow \eta\eta^\prime) = {g_{\eta\eta^\prime}^2 \over 24 \pi m_{\eta_1}^2 } |\vec q_\eta|^3 \, .
\end{equation}
where $\vec q_\eta$ is the three-momentum of the final state $\eta$. Numerically, we obtain
\begin{eqnarray}
g_{\eta\eta^\prime} &=& 3.08^{+1.30}_{-0.91} \  \, ,
\\
\Gamma(\eta_1 \xrightarrow{c} \eta\eta^\prime) &=& {5.0^{+4.6}_{-3.1}~\rm MeV} \, .
\end{eqnarray}

Similarly, we study the other normal and abnormal decay processes. The obtained results are summarized in Table~\ref{tab:decay}.

\begin{table*}[tbh]
\begin{center}
\renewcommand{\arraystretch}{1.5}
\caption{Partial decay widths of the hybrid states $| \bar q q g; 1^-1^{-+} \rangle$, $| \bar q q g; 0^+1^{-+} \rangle$, and $| \bar s s g; 0^+1^{-+} \rangle$, in units of MeV. $\Gamma(\pi_1/\eta_1 \xrightarrow{a,b,c} M_1 M_2)$ are related to the processes depicted in Figs.~\ref{fig:decay}(a,b,c), respectively. We simply sum over the partial decay widths to obtain the total decay widths, as listed in the last row. Note that the decay channels $\pi_1 \xrightarrow{b,c} \eta \pi/\eta^\prime \pi$ and $\pi_1/\eta_1 \rightarrow K^*(892)\bar K / K_1(1270)\bar K / K^*(892) \bar K^*(892)$ have not been investigated in our previous QCD sum rule studies~\cite{Chen:2010ic,Huang:2010dc,Chen:2022qpd}.}
\begin{tabular}{c|c|c|c}
\hline \hline
\multirow{2}*{~~~~~~~~~~~~~~Channel~~~~~~~~~~~~~~}       & ~~~~~~~$| \bar q q g; 1^-1^{-+} \rangle$~~~~~~~ & ~~~~~~~$| \bar q q g; 0^+1^{-+} \rangle$~~~~~~~ & ~~~~~~~$| \bar s s g; 0^+1^{-+} \rangle$~~~~~~~
\\                                                       & $M=1.67^{+0.15}_{-0.17}$~GeV                    & $M=1.67^{+0.15}_{-0.17}$~GeV                    & $M=1.84^{+0.14}_{-0.15}$~GeV
\\ \hline \hline
$\pi_1/\eta_1 \rightarrow \rho \pi$                      & $242^{+310}_{-179}$                             & --                                              & --
\\ \hline
$\pi_1/\eta_1 \rightarrow b_1(1235) \pi$                 & $14.5^{+25.9}_{-13.9}$                           & --                                              & --
\\ \hline
$\pi_1/\eta_1 \rightarrow f_1(1285) \pi$                 & $35.9^{+53.9}_{-36.4}$                          & --                                              & --
\\ \hline
$\pi_1/\eta_1 \rightarrow \eta \pi$                      & $2.3^{+2.5}_{-1.2}$                             & --                                              & --
\\ \hline
$\pi_1/\eta_1 \xrightarrow{b} \eta \pi$                  & $57.8^{+65.0}_{-31.4}$                          & --                                              & --
\\ \hline
$\pi_1/\eta_1 \rightarrow \eta^\prime \pi$               & $0.43^{+0.50}_{-0.28}$                          & --                                              & --
\\ \hline
$\pi_1/\eta_1 \xrightarrow{c} \eta^\prime \pi$           & $149^{+162}_{-~78}$                             & --                                              & --
\\ \hline
$\pi_1/\eta_1 \rightarrow a_1(1260) \pi$                 & --                                              & $79.5^{+112.4}_{-~74.9}$                          & --
\\ \hline
$\pi_1/\eta_1 \xrightarrow{a} \eta \eta^\prime$          & --                                              & $0.07^{+0.12}_{-0.07}$                          & $0.93^{+1.04}_{-0.69}$
\\ \hline
$\pi_1/\eta_1 \xrightarrow{b} \eta \eta^\prime$          & --                                              & $1.62^{+2.13}_{-1.61}$                          & $1.64^{+1.51}_{-1.01}$
\\ \hline
$\pi_1/\eta_1 \xrightarrow{c} \eta \eta^\prime$          & --                                              & $11.5^{+11.7}_{-11.5}$                          & $5.0^{+4.6}_{-3.1}$
\\ \hline
$\pi_1/\eta_1 \rightarrow K^*(892)\bar K+c.c.$           & $25.3^{+34.7}_{-24.7}$                          & $25.3^{+34.7}_{-24.7}$                          & $73.9^{+85.7}_{-58.0}$
\\ \hline
$\pi_1/\eta_1 \rightarrow K_1(1270)\bar K+c.c.$          & --                                              & --                                              & $14.6^{+19.8}_{-14.6}$
\\ \hline
$\pi_1/\eta_1 \rightarrow K^*(892) \bar K^*(892)$          & --                                              & --                                              & $0.08^{+0.39}_{-0.08}$
\\ \hline \hline
Sum                                                      & $530^{+540}_{-330}$                             & $120^{+160}_{-110}$                             & $100^{+110}_{-~80}$
\\ \hline \hline
\end{tabular}
\label{tab:decay}
\end{center}
\end{table*}

%
\section{Summary and Discussions}
\label{sec:summary}

In this paper we study the single-gluon hybrid states with various (exotic) quantum numbers. We systematically construct twenty-four single-gluon hybrid currents, and use eighteen of them to perform QCD sum rule analyses. We calculate the masses of forty-four single-gluon hybrid states with the quark-gluon contents $\bar q q g$ ($q=u/d$) and $\bar s s g$. The obtained results are summarized in Table~\ref{tab:mass1} and Table~\ref{tab:mass2}. Especially, the masses and decay constants of the $J^{PC} = 1^{-+}$ hybrid states are extracted from the current $J^{\mu}_{1^{-+}}$ given in Eq.~(\ref{def:A1mp}) to be:
\begin{eqnarray}
\nonumber M_{|\bar q q g;1^{-+}\rangle} &=& 1.67^{+0.15}_{-0.17}{\rm~GeV} \, ,
\\ \nonumber f_{|\bar q q g;1^{-+}\rangle} &=& 0.243^{+0.057}_{-0.052}{\rm~GeV}^4 \, ,
\\ \nonumber M_{|\bar s s g;1^{-+}\rangle} &=& 1.84^{+0.14}_{-0.15}{\rm~GeV} \, ,
\\ \nonumber f_{|\bar s s g;1^{-+}\rangle} &=& 0.300^{+0.063}_{-0.058}{\rm~GeV}^4 \, .
\end{eqnarray}
Since we do not differentiate the up and down quarks within the QCD sum rule method, the masses and decay constants of the states in the same isospin multiplet are calculated to be the same:
\begin{eqnarray}
\nonumber M_{|\bar q q g;1^-1^{-+}\rangle} = M_{|\bar q q g;0^+1^{-+}\rangle} &=&1.67^{+0.15}_{-0.17}{\rm~GeV} \, ,
\\ \nonumber f_{|\bar q q g;1^-1^{-+}\rangle} = f_{|\bar q q g;0^+1^{-+}\rangle} &=& 0.243^{+0.057}_{-0.052}{\rm~GeV}^4 \, ,
\\ \nonumber M_{|\bar s s g;0^+1^{-+}\rangle} &=& 1.84^{+0.14}_{-0.15}{\rm~GeV} \, ,
\\ \nonumber f_{|\bar s s g;0^+1^{-+}\rangle} &=& 0.300^{+0.063}_{-0.058}{\rm~GeV}^4 \, .
\end{eqnarray}
There have been a lot of Lattice QCD calculations on the $I^GJ^{PC} = 1^-1^{-+}$ hybrid state in the past fifty years~\cite{Griffiths:1983ah,Michael:1985ne,Perantonis:1990dy,Foster:1998wu,Juge:2002br,Bali:2003jq}, and especially, the Hadron Spectrum collaboration have performed exhaustive analyses~\cite{Dudek:2009qf,Dudek:2010wm,Dudek:2011bn,HadronSpectrum:2012gic,Dudek:2013yja,Woss:2020ayi,Briceno:2017max}. C.~A.~Meyer and E.~S.~Swanson summarized these results, and the naive extrapolation of its mass, calculated by Lattice QCD, to the physical pion mass turns out to be approximately 1.6~GeV~\cite{Meyer:2015eta}. Therefore, our QCD sum rule calculation is well consistent with the Lattice QCD calculations.

Based on the mass calculations, we systematically study the decay properties of the $J^{PC} = 1^{-+}$ hybrid states. We have considered the normal decay process depicted in Fig.~\ref{fig:decay}(a). We have also considered the abnormal decay processes depicted in Figs.~\ref{fig:decay}(b,c), with the $\eta^{(\prime)}$ mesons produced by the QCD axial anomaly. The obtained results are summarized in Table~\ref{tab:decay}, and especially,
\begin{eqnarray}
\nonumber \Gamma_{|\bar q q g;1^-1^{-+}\rangle} &=& 530^{+540}_{-330}{\rm~MeV} \, ,
\\ \nonumber \Gamma_{|\bar q q g;0^+1^{-+}\rangle} &=& 120^{+160}_{-110}{\rm~MeV} \, ,
\\ \nonumber \Gamma_{|\bar s s g;0^+1^{-+}\rangle} &=& 100^{+110}_{-~80}{\rm~MeV} \, .
\end{eqnarray}

The above QCD sum rule results suggest that the $\pi_1(1600)$ and $\eta_1(1855)$ can be respectively interpreted as the single-gluon hybrid states $|\bar q q g;1^-1^{-+}\rangle$ and $|\bar s s g;0^+1^{-+}\rangle$, so there exists another isoscalar state $|\bar q q g;0^+1^{-+}\rangle$, whose mass and width are smaller than those of $\eta_1(1855)$. Considering the uncertainties, our results suggest that the $\pi_1(1600)$ and $\eta_1(1855)$ may also be respectively interpreted as the single-gluon hybrid states $|\bar q q g;1^-1^{-+}\rangle$ and $|\bar q q g;0^+1^{-+}\rangle$, so there exists another isoscalar state $|\bar s s g;0^+1^{-+}\rangle$, whose mass and width are larger than those of $\eta_1(1855)$. To differentiate these two assignments, it is useful to examine the $a_1(1260) \pi$ decay channel. We find in Table~\ref{tab:decay} that the $\eta^{(\prime)}$-relevant decay modes, as the characteristic decay modes of hybrid states, are enhanced to some extent. To verify whether the exotic $\pi_1$ and $\eta_1$ resonances are hybrid states or not, we propose to examine these decay modes in future BESIII, Belle-II, GlueX, LHCb, and PANDA experiments.

%
\section*{Acknowledgments}
%

This project is supported by
the National Natural Science Foundation of China under Grant No.~12075019,
the Jiangsu Provincial Double-Innovation Program under Grant No.~JSSCRC2021488,
and
the Fundamental Research Funds for the Central Universities.

\appendix
\section{Spectral densities}
\label{app:ope}

In this appendix we show the OPE spectral densities extracted from the single-gluon hybrid currents given in Eqs.~(\ref{def:B1mm}-\ref{def:At1mm}) and Eqs.~(\ref{def:A0pp}-\ref{def:Bt2pp}) with the quark-gluon content $\bar s s g$. Those with the quark-gluon content $\bar q q g$ ($q=u/d$) can be easily derived by replacing $m_s \rightarrow 0$, $\left\langle {\bar ss} \right\rangle \rightarrow \left\langle {\bar qq} \right\rangle$, and $\left\langle {{g_s}\bar s\sigma Gs} \right\rangle \rightarrow \left\langle {{g_s}\bar q\sigma Gq} \right\rangle$.
\begin{widetext}
\begin{align}
    \Pi^{\alpha\beta}_{1^{--}} {\left( {M_B^2,{s_0}} \right)} & = \int_{4m_s^2}^{{s_0}} {(\frac{{{s^3}{\alpha _s}}}{{240{\pi ^3}}}}  - \frac{{m_s^2{s^2}{\alpha _s}}}{{24{\pi ^3}}} + s(\frac{{\left\langle {{\alpha _s}GG} \right\rangle }}{{48{\pi ^2}}} - \frac{{\left\langle {{\alpha _s}GG} \right\rangle {\alpha _s}}}{{1152{\pi ^3}}} - \frac{{2{m_s}\left\langle {\bar ss} \right\rangle {\alpha _s}}}{{9\pi }})
    \\ \nonumber
    & - \frac{{\left\langle {g_s^3{G^3}} \right\rangle }}{{96{\pi ^2}}} - \frac{{\left\langle {{\alpha _s}GG} \right\rangle m_s^2}}{{24{\pi ^2}}})\times {e^{ - s/M_B^2}}ds+ ( - \frac{{{{\left\langle {{\alpha _s}GG} \right\rangle }^2}}}{{4608{\pi ^2}}} - \frac{{\left\langle {{\alpha _s}GG} \right\rangle {m_s}\left\langle {\bar ss} \right\rangle }}{{18}}
    \\ \nonumber
    & - \frac{4}{9}m_s^2\pi {\left\langle {\bar ss} \right\rangle ^2}{\alpha _s} + \frac{8}{9}\pi \left\langle {\bar ss} \right\rangle \left\langle {{g_s}\bar s\sigma Gs} \right\rangle {\alpha _s}) \, ,
	\\
	\widetilde{\Pi}^{\alpha\beta}_{1^{+-}}{\left( {M_B^2,{s_0}} \right)}&= \int_{4m_s^2}^{{s_0}} {(\frac{{{s^3}{\alpha _s}}}{{60{\pi ^3}}}}  - \frac{{m_s^2{s^2}{\alpha _s}}}{{6{\pi ^3}}} + s( - \frac{{\left\langle {{\alpha _s}GG} \right\rangle }}{{12{\pi ^2}}} - \frac{{\left\langle {{\alpha _s}GG} \right\rangle {\alpha _s}}}{{288{\pi ^3}}} - \frac{{8{m_s}\left\langle {\bar ss} \right\rangle {\alpha _s}}}{{9\pi }})
	\\ \nonumber
	&  + \frac{{\left\langle {g_s^3{G^3}} \right\rangle }}{{24{\pi ^2}}} + \frac{{\left\langle {{\alpha _s}GG} \right\rangle m_s^2}}{{6{\pi ^2}}})\times {e^{ - s/M_B^2}}ds+ (\frac{{{{\left\langle {{\alpha _s}GG} \right\rangle }^2}}}{{1152{\pi ^2}}} - \frac{{\left\langle {g_s^3{G^3}} \right\rangle m_s^2}}{{12{\pi ^2}}}
	\\ \nonumber
	&+ \frac{2}{9}\left\langle {{\alpha _s}GG} \right\rangle {m_s}\left\langle {\bar ss} \right\rangle  - \frac{{16}}{9}\pi m_s^2{\left\langle {\bar ss} \right\rangle ^2}{\alpha _s} + \frac{{32}}{9}\pi \left\langle {\bar ss} \right\rangle \left\langle {{g_s}\bar s\sigma Gs} \right\rangle {\alpha _s}) \, ,
	\\
    \Pi^{\alpha\beta}_{1^{+-}} {\left( {M_B^2,{s_0}} \right)}& = \int_{4m_s^2}^{{s_0}} {(\frac{{{s^3}{\alpha _s}}}{{240{\pi ^3}}}}  - \frac{{m_s^2{s^2}{\alpha _s}}}{{8{\pi ^3}}} + s(\frac{{\left\langle {{\alpha _s}GG} \right\rangle }}{{48{\pi ^2}}} - \frac{{\left\langle {{\alpha _s}GG} \right\rangle {\alpha _s}}}{{1152{\pi ^3}}} + \frac{{2{m_s}\left\langle {\bar ss} \right\rangle {\alpha _s}}}{{3\pi }})
	\\ \nonumber
	&- \frac{{\left\langle {g_s^3{G^3}} \right\rangle }}{{96{\pi ^2}}} - \frac{{\left\langle {{\alpha _s}GG} \right\rangle m_s^2}}{{8{\pi ^2}}})\times {e^{ - s/M_B^2}}ds+ ( - \frac{{{{\left\langle {{\alpha _s}GG} \right\rangle }^2}}}{{4608{\pi ^2}}} + \frac{{\left\langle {{\alpha _s}GG} \right\rangle {m_s}\left\langle {\bar ss} \right\rangle }}{6}
	\\ \nonumber
	&- \frac{4}{9}m_s^2\pi {\left\langle {\bar ss} \right\rangle ^2}{\alpha _s} - \frac{8}{9}\pi \left\langle {\bar ss} \right\rangle \left\langle {{g_s}\bar s\sigma Gs} \right\rangle {\alpha _s}) \, ,
	\\
	\widetilde{\Pi}^{\alpha\beta}_{1^{--}}{\left( {M_B^2,{s_0}} \right)}& = \int_{4m_s^2}^{{s_0}} {(\frac{{{s^3}{\alpha _s}}}{{60{\pi ^3}}}}  - \frac{{m_s^2{s^2}{\alpha _s}}}{{2{\pi ^3}}} + s( - \frac{{\left\langle {{\alpha _s}GG} \right\rangle }}{{12{\pi ^2}}} - \frac{{\left\langle {{\alpha _s}GG} \right\rangle {\alpha _s}}}{{288{\pi ^3}}} + \frac{{8{m_s}\left\langle {\bar ss} \right\rangle {\alpha _s}}}{{3\pi }})
	\\ \nonumber
	& + \frac{{\left\langle {g_s^3{G^3}} \right\rangle }}{{24{\pi ^2}}} + \frac{{\left\langle {{\alpha _s}GG} \right\rangle m_s^2}}{{2{\pi ^2}}})\times {e^{ - s/M_B^2}}ds+ (\frac{{{{\left\langle {{\alpha _s}GG} \right\rangle }^2}}}{{1152{\pi ^2}}} - \frac{{\left\langle {g_s^3{G^3}} \right\rangle m_s^2}}{{4{\pi ^2}}}
	\\ \nonumber
	&- \frac{2}{3}\left\langle {{\alpha _s}GG} \right\rangle {m_s}\left\langle {\bar ss} \right\rangle  - \frac{{16}}{9}\pi m_s^2{\left\langle {\bar ss} \right\rangle ^2}{\alpha _s} - \frac{{32}}{9}\pi \left\langle {\bar ss} \right\rangle \left\langle {{g_s}\bar s\sigma Gs} \right\rangle {\alpha _s}) \, ,
	\\
	\Pi^{\mu}_{1^{-+}} {\left( {M_B^2,{s_0}} \right)} &= \int_{4m_s^2}^{{s_0}} {(\frac{{{s^3}{\alpha _s}}}{{60{\pi ^3}}}}  - \frac{{m_s^2{s^2}{\alpha _s}}}{{3{\pi ^3}}} + s(\frac{{\left\langle {{\alpha _s}GG} \right\rangle }}{{36{\pi ^2}}} + \frac{{13\left\langle {{\alpha _s}GG} \right\rangle {\alpha _s}}}{{432{\pi ^3}}} + \frac{{8{m_s}\left\langle {\bar ss} \right\rangle {\alpha _s}}}{{9\pi }})
	\\ \nonumber
	&+ \frac{{\left\langle {g_s^3{G^3}} \right\rangle }}{{32{\pi ^2}}} - \frac{{3\left\langle {{\alpha _s}GG} \right\rangle m_s^2{\alpha _s}}}{{64{\pi ^3}}} - \frac{{3{m_s}\left\langle {{g_s}\bar s\sigma Gs} \right\rangle {\alpha _s}}}{{4\pi }}) \times {e^{ - s/M_B^2}}ds
	\\ \nonumber
	&+ (\frac{{{{\left\langle {{\alpha _s}GG} \right\rangle }^2}}}{{3456{\pi ^2}}} - \frac{{\left\langle {g_s^3{G^3}} \right\rangle m_s^2}}{{16{\pi ^2}}} - \frac{2}{9}\left\langle {{\alpha _s}GG} \right\rangle {m_s}\left\langle {\bar ss} \right\rangle  + \frac{{11}}{9}\pi \left\langle {\bar ss} \right\rangle \left\langle {{g_s}\bar s\sigma Gs} \right\rangle {\alpha _s}) \, ,
	\\
	\widetilde{\Pi}^{\mu}_{1^{++}}{\left( {M_B^2,{s_0}} \right)}&= \int_{4m_s^2}^{{s_0}} {(\frac{{{s^3}{\alpha _s}}}{{15{\pi ^3}}}}  - \frac{{4m_s^2{s^2}{\alpha _s}}}{{3{\pi ^3}}} + s( - \frac{{\left\langle {{\alpha _s}GG} \right\rangle }}{{9{\pi ^2}}} - \frac{{5\left\langle {{\alpha _s}GG} \right\rangle {\alpha _s}}}{{108{\pi ^3}}} + \frac{{32{m_s}\left\langle {\bar ss} \right\rangle {\alpha _s}}}{{9\pi }})
    \\ \nonumber
	&- \frac{{\left\langle {g_s^3{G^3}} \right\rangle }}{{4{\pi ^2}}} + \frac{{3\left\langle {{\alpha _s}GG} \right\rangle m_s^2{\alpha _s}}}{{16{\pi ^3}}} + \frac{{3{m_s}\left\langle {{g_s}\bar s\sigma Gs} \right\rangle {\alpha _s}}}{\pi }) \times {e^{ - s/M_B^2}}ds+ ( - \frac{{{{\left\langle {{\alpha _s}GG} \right\rangle }^2}}}{{864{\pi ^2}}}
	\\ \nonumber
	&+ \frac{{\left\langle {g_s^3{G^3}} \right\rangle m_s^2}}{{3{\pi ^2}}} + \frac{8}{9}\left\langle {{\alpha _s}GG} \right\rangle {m_s}\left\langle {\bar ss} \right\rangle  - \frac{{\left\langle {{\alpha _s}GG} \right\rangle {m_s}\left\langle {\bar ss} \right\rangle {\alpha _s}}}{{3\pi }} - \frac{{28}}{9}\pi \left\langle {\bar ss} \right\rangle \left\langle {{g_s}\bar s\sigma Gs} \right\rangle {\alpha _s}) \, ,
	\\
	\Pi^{\mu}_{1^{+-}} {\left( {M_B^2,{s_0}} \right)} &= \int_{4m_s^2}^{{s_0}} {(\frac{{{s^3}{\alpha _s}}}{{60{\pi ^3}}}}  - \frac{{5m_s^2{s^2}{\alpha _s}}}{{12{\pi ^3}}} + s(\frac{{\left\langle {{\alpha _s}GG} \right\rangle }}{{36{\pi ^2}}} + \frac{{13\left\langle {{\alpha _s}GG} \right\rangle {\alpha _s}}}{{432{\pi ^3}}} + \frac{{16{m_s}\left\langle {\bar ss} \right\rangle {\alpha _s}}}{{9\pi }})
	\\ \nonumber
	&+ \frac{{\left\langle {g_s^3{G^3}} \right\rangle }}{{32{\pi ^2}}} - \frac{{\left\langle {{\alpha _s}GG} \right\rangle m_s^2}}{{4{\pi ^2}}} - \frac{{3\left\langle {{\alpha _s}GG} \right\rangle m_s^2{\alpha _s}}}{{64{\pi ^3}}} + \frac{{3{m_s}\left\langle {{g_s}\bar s\sigma Gs} \right\rangle {\alpha _s}}}{{4\pi }}) \times {e^{ - s/M_B^2}}ds
	\\ \nonumber
	&+ (\frac{{{{\left\langle {{\alpha _s}GG} \right\rangle }^2}}}{{3456{\pi ^2}}} - \frac{{\left\langle {g_s^3{G^3}} \right\rangle m_s^2}}{{48{\pi ^2}}} + \frac{4}{9}\left\langle {{\alpha _s}GG} \right\rangle {m_s}\left\langle {\bar ss} \right\rangle  - \frac{{11}}{9}\pi \left\langle {\bar ss} \right\rangle \left\langle {{g_s}\bar s\sigma Gs} \right\rangle {\alpha _s}) \, ,
	\\
	\widetilde{\Pi}^{\mu}_{1^{--}}{\left( {M_B^2,{s_0}} \right)} &= \int_{4m_s^2}^{{s_0}} {(\frac{{{s^3}{\alpha _s}}}{{15{\pi ^3}}}}  - \frac{{5m_s^2{s^2}{\alpha _s}}}{{3{\pi ^3}}} + s( - \frac{{\left\langle {{\alpha _s}GG} \right\rangle }}{{9{\pi ^2}}} - \frac{{5\left\langle {{\alpha _s}GG} \right\rangle {\alpha _s}}}{{108{\pi ^3}}} + \frac{{64{m_s}\left\langle {\bar ss} \right\rangle {\alpha _s}}}{{9\pi }})
	\\ \nonumber
	&- \frac{{\left\langle {g_s^3{G^3}} \right\rangle }}{{4{\pi ^2}}} + \frac{{\left\langle {{\alpha _s}GG} \right\rangle m_s^2}}{{{\pi ^2}}} + \frac{{3\left\langle {{\alpha _s}GG} \right\rangle m_s^2{\alpha _s}}}{{16{\pi ^3}}} - \frac{{3{m_s}\left\langle {{g_s}\bar s\sigma Gs} \right\rangle {\alpha _s}}}{\pi }) \times {e^{ - s/M_B^2}}ds
	\\ \nonumber
	&+ ( - \frac{{{{\left\langle {{\alpha _s}GG} \right\rangle }^2}}}{{864{\pi ^2}}} - \frac{{16}}{9}\left\langle {{\alpha _s}GG} \right\rangle {m_s}\left\langle {\bar ss} \right\rangle  - \frac{{\left\langle {{\alpha _s}GG} \right\rangle {m_s}\left\langle {\bar ss} \right\rangle {\alpha _s}}}{{3\pi }} + \frac{{28}}{9}\pi \left\langle {\bar ss} \right\rangle \left\langle {{g_s}\bar s\sigma Gs} \right\rangle {\alpha _s}) \, ,
	\\
	\Pi^{\mu}_{0^{++}} {\left( {M_B^2,{s_0}} \right)} &= \int_{4m_s^2}^{{s_0}} {(\frac{{{s^3}{\alpha _s}}}{{120{\pi ^3}}}}  +s(- \frac{{\left\langle {{\alpha _s}GG} \right\rangle }}{{24{\pi ^2}}} + \frac{{\left\langle {{\alpha _s}GG} \right\rangle {\alpha _s}}}{{576{\pi ^3}}} - \frac{{4{m_s}\left\langle {\bar ss} \right\rangle {\alpha _s}}}{{3\pi }}) - \frac{{3\left\langle {g_s^3{G^3}} \right\rangle }}{{32{\pi ^2}}}
	\\ \nonumber
	& + \frac{{\left\langle {{\alpha _s}GG} \right\rangle m_s^2}}{{12{\pi ^2}}}- \frac{{3\left\langle {{\alpha _s}GG} \right\rangle m_s^2{\alpha _s}}}{{64{\pi ^3}}} + \frac{{32}}{9}\pi {\left\langle {\bar ss} \right\rangle ^2}{\alpha _s} + \frac{{13\left\langle {{g_s}\bar s\sigma Gs} \right\rangle {m_s}{\alpha _s}}}{{12\pi }}) \times {e^{ - s/M_B^2}}ds
	\\ \nonumber
	& +( - \frac{{{{\left\langle {{\alpha _s}GG} \right\rangle }^2}}}{{2304{\pi ^2}}} - \frac{{\left\langle {g_s^3{G^3}} \right\rangle m_s^2}}{{16{\pi ^2}}} + \frac{1}{3}\left\langle {{\alpha _s}GG} \right\rangle {m_s}\left\langle {\bar ss} \right\rangle  - \frac{{\left\langle {{\alpha _s}GG} \right\rangle {m_s}\left\langle {\bar ss} \right\rangle {\alpha _s}}}{{8\pi }}
	\\ \nonumber
	&-\frac{{8\pi m_s^2{{\left\langle {\bar ss} \right\rangle }^2}{\alpha _s}}}{3} - \frac{1}{3}\pi \left\langle {\bar ss} \right\rangle \left\langle {{g_s}\bar s\sigma Gs} \right\rangle {\alpha _s}) \, ,
	\\
	\widetilde{\Pi}^{\mu}_{0^{-+}}{\left( {M_B^2,{s_0}} \right)}&= \int_{4m_s^2}^{{s_0}} {(\frac{{{s^3}{\alpha _s}}}{{30{\pi ^3}}}}  + s(\frac{{\left\langle {{\alpha _s}GG} \right\rangle }}{{6{\pi ^2}}} + \frac{{37\left\langle {{\alpha _s}GG} \right\rangle {\alpha _s}}}{{144{\pi ^3}}} - \frac{{16{m_s}\left\langle {\bar ss} \right\rangle {\alpha _s}}}{{3\pi }})+\frac{{\left\langle {g_s^3{G^3}} \right\rangle }}{{4{\pi ^2}}} - \frac{{\left\langle {{\alpha _s}GG} \right\rangle m_s^2}}{{3{\pi ^2}}}
	\\ \nonumber
    &- \frac{{13\left\langle {{\alpha _s}GG} \right\rangle m_s^2{\alpha _s}}}{{16{\pi ^3}}} + \frac{{128}}{9}\pi {\left\langle {\bar ss} \right\rangle ^2}{\alpha _s} - \frac{{5\left\langle {{g_s}\bar s\sigma Gs} \right\rangle {m_s}{\alpha _s}}}{{3\pi }}) \times {e^{ - s/M_B^2}}ds
	\\ \nonumber
	&+(\frac{{{{\left\langle {{\alpha _s}GG} \right\rangle }^2}}}{{576{\pi ^2}}}- \frac{{\left\langle {g_s^3{G^3}} \right\rangle m_s^2}}{{2{\pi ^2}}} - \frac{4}{3}\left\langle {{\alpha _s}GG} \right\rangle {m_s}\left\langle {\bar ss} \right\rangle  - \frac{{\left\langle {{\alpha _s}GG} \right\rangle {m_s}\left\langle {\bar ss} \right\rangle {\alpha _s}}}{{2\pi }}
	\\ \nonumber
	&- \frac{{32\pi m_s^2{{\left\langle {\bar ss} \right\rangle }^2}{\alpha _s}}}{3} + \frac{{20}}{3}\pi \left\langle {\bar ss} \right\rangle \left\langle {{g_s}\bar s\sigma Gs} \right\rangle {\alpha _s}) \, ,
	\\
    \Pi^{\mu}_{0^{--}} {\left( {M_B^2,{s_0}} \right)} &= \int_{4m_s^2}^{{s_0}} {(\frac{{{s^3}{\alpha _s}}}{{120{\pi ^3}}}}  - \frac{{m_s^2s^2{\alpha _s}}}{{4{\pi ^3}}} +s(- \frac{{\left\langle {{\alpha _s}GG} \right\rangle }}{{24{\pi ^2}}} + \frac{{\left\langle {{\alpha _s}GG} \right\rangle {\alpha _s}}}{{576{\pi ^3}}} + \frac{{4{m_s}\left\langle {\bar ss} \right\rangle {\alpha _s}}}{{3\pi }}) - \frac{{3\left\langle {g_s^3{G^3}} \right\rangle }}{{32{\pi ^2}}}
	\\ \nonumber
	& + \frac{{\left\langle {{\alpha _s}GG} \right\rangle m_s^2}}{{3{\pi ^2}}} - \frac{{3\left\langle {{\alpha _s}GG} \right\rangle m_s^2{\alpha _s}}}{{64{\pi ^3}}} - \frac{{32}}{9}\pi {\left\langle {\bar ss} \right\rangle ^2}{\alpha _s} - \frac{{13\left\langle {{g_s}\bar s\sigma Gs} \right\rangle {m_s}{\alpha _s}}}{{12\pi }})\times {e^{ - s/M_B^2}}ds
	\\ \nonumber
	&+ ( - \frac{{{{\left\langle {{\alpha _s}GG} \right\rangle }^2}}}{{2304{\pi ^2}}} - \frac{{3\left\langle {g_s^3{G^3}} \right\rangle m_s^2}}{{16{\pi ^2}}} - \frac{1}{3}\left\langle {{\alpha _s}GG} \right\rangle {m_s}\left\langle {\bar ss} \right\rangle  - \frac{{\left\langle {{\alpha _s}GG} \right\rangle {m_s}\left\langle {\bar ss} \right\rangle {\alpha _s}}}{{8\pi }}
	\\ \nonumber
	 &- \frac{{8\pi m_s^2{{\left\langle {\bar ss} \right\rangle }^2}{\alpha _s}}}{3} + \frac{1}{3}\pi \left\langle {\bar ss} \right\rangle \left\langle {{g_s}\bar s\sigma Gs} \right\rangle {\alpha _s}) \, ,
	\\
	\widetilde{\Pi}^{\mu}_{0^{+-}}{\left( {M_B^2,{s_0}} \right)} &= \int_{4m_s^2}^{{s_0}} {(\frac{{{s^3}{\alpha _s}}}{{30{\pi ^3}}}}  - \frac{{m_s^2s^2{\alpha _s}}}{{{\pi ^3}}} +s(+ \frac{{\left\langle {{\alpha _s}GG} \right\rangle }}{{6{\pi ^2}}} + \frac{{37\left\langle {{\alpha _s}GG} \right\rangle {\alpha _s}}}{{144{\pi ^3}}} + \frac{{16{m_s}\left\langle {\bar ss} \right\rangle {\alpha _s}}}{{3\pi }}) +\frac{{\left\langle {g_s^3{G^3}} \right\rangle }}{{4{\pi ^2}}}
	\\ \nonumber
	& - \frac{{4\left\langle {{\alpha _s}GG} \right\rangle m_s^2}}{{3{\pi ^2}}} - \frac{{13\left\langle {{\alpha _s}GG} \right\rangle m_s^2{\alpha _s}}}{{16{\pi ^3}}} - \frac{{128}}{9}\pi {\left\langle {\bar ss} \right\rangle ^2}{\alpha _s} + \frac{{5\left\langle {{g_s}\bar s\sigma Gs} \right\rangle {m_s}{\alpha _s}}}{{3\pi }})\times {e^{ - s/M_B^2}}ds
	\\ \nonumber
	&+ (\frac{{{{\left\langle {{\alpha _s}GG} \right\rangle }^2}}}{{576{\pi ^2}}} - \frac{{\left\langle {g_s^3{G^3}} \right\rangle m_s^2}}{{2{\pi ^2}}} + \frac{4}{3}\left\langle {{\alpha _s}GG} \right\rangle {m_s}\left\langle {\bar ss} \right\rangle  - \frac{{\left\langle {{\alpha _s}GG} \right\rangle {m_s}\left\langle {\bar ss} \right\rangle {\alpha _s}}}{{2\pi }}
	\\ \nonumber
	&- \frac{{32\pi m_s^2{{\left\langle {\bar ss} \right\rangle }^2}{\alpha _s}}}{3} - \frac{{20}}{3}\pi \left\langle {\bar ss} \right\rangle \left\langle {{g_s}\bar s\sigma Gs} \right\rangle {\alpha _s}) \, ,
	\\
	\Pi_{0^{++}} {\left( {M_B^2,{s_0}} \right)}  &= \int_{4m_s^2}^{{s_0}} {(\frac{{{s^3}{\alpha _s}}}{{6{\pi ^3}}}}  - \frac{{4m_s^2{s^2}{\alpha _s}}}{{{\pi ^3}}} + s(\frac{{37\left\langle {{\alpha _s}GG} \right\rangle {\alpha _s}}}{{48{\pi ^3}}} + \frac{{16{m_s}\left\langle {\bar ss} \right\rangle {\alpha _s}}}{\pi })- \frac{{\left\langle {g_s^3{G^3}} \right\rangle }}{{4{\pi ^2}}}
	\\ \nonumber
	& - \frac{{\left\langle {{\alpha _s}GG} \right\rangle m_s^2}}{{{\pi ^2}}} - \frac{{3\left\langle {{\alpha _s}GG} \right\rangle m_s^2{\alpha _s}}}{{2{\pi ^3}}} + \frac{{12{m_s}\left\langle {{g_s}\bar s\sigma Gs} \right\rangle {\alpha _s}}}{\pi }) \times {e^{ - s/M_B^2}}ds+ (\frac{{3\left\langle {g_s^3{G^3}} \right\rangle m_s^2}}{{4{\pi ^2}}}
	\\ \nonumber
	&+ \frac{8}{3}\left\langle {{\alpha _s}GG} \right\rangle {m_s}\left\langle {\bar ss} \right\rangle  + \frac{{\left\langle {{\alpha _s}GG} \right\rangle {m_s}\left\langle {\bar ss} \right\rangle {\alpha _s}}}{\pi } + \frac{{32\pi m_s^2{{\left\langle {\bar ss} \right\rangle }^2}{\alpha _s}}}{3} - 16\pi \left\langle {\bar ss} \right\rangle \left\langle {{g_s}\bar s\sigma Gs} \right\rangle {\alpha _s}) \, ,
	\\
	\Pi_{0^{-+}} {\left( {M_B^2,{s_0}} \right)} &= \int_{4m_s^2}^{{s_0}} {(\frac{{{s^3}{\alpha _s}}}{{6{\pi ^3}}}}  - \frac{{4m_s^2{s^2}{\alpha _s}}}{{{\pi ^3}}} + s(\frac{{37\left\langle {{\alpha _s}GG} \right\rangle {\alpha _s}}}{{48{\pi ^3}}} + \frac{{16{m_s}\left\langle {\bar ss} \right\rangle {\alpha _s}}}{\pi })- \frac{{\left\langle {g_s^3{G^3}} \right\rangle }}{{4{\pi ^2}}}+ \frac{{\left\langle {{\alpha _s}GG} \right\rangle m_s^2}}{{{\pi ^2}}}
	\\ \nonumber
	& - \frac{{3\left\langle {{\alpha _s}GG} \right\rangle m_s^2{\alpha _s}}}{{2{\pi ^3}}} - \frac{{12{m_s}\left\langle {{g_s}\bar s\sigma Gs} \right\rangle {\alpha _s}}}{\pi }) \times {e^{ - s/M_B^2}}ds+ (\frac{{\left\langle {g_s^3{G^3}} \right\rangle m_s^2}}{{4{\pi ^2}}} - \frac{8}{3}\left\langle {{\alpha _s}GG} \right\rangle {m_s}\left\langle {\bar ss} \right\rangle
	\\ \nonumber
	&  + \frac{{\left\langle {{\alpha _s}GG} \right\rangle {m_s}\left\langle {\bar ss} \right\rangle {\alpha _s}}}{\pi } + \frac{{32\pi m_s^2{{\left\langle {\bar ss} \right\rangle }^2}{\alpha _s}}}{3} + 16\pi \left\langle {\bar ss} \right\rangle \left\langle {{g_s}\bar s\sigma Gs} \right\rangle {\alpha _s}) \, ,
	\\
	\Pi^{\alpha\beta}_{1^{++}} {\left( {M_B^2,{s_0}} \right)} &= \int_{4m_s^2}^{{s_0}} {(\frac{{{s^3}{\alpha _s}}}{{60{\pi ^3}}}}  - \frac{{m_s^2{s^2}{\alpha _s}}}{{3{\pi ^3}}} + s(\frac{{19\left\langle {{\alpha _s}GG} \right\rangle {\alpha _s}}}{{864{\pi ^3}}} + \frac{{8{m_s}\left\langle {\bar ss} \right\rangle {\alpha _s}}}{{9\pi }}) - \frac{{\left\langle {{\alpha _s}GG} \right\rangle m_s^2}}{{6{\pi ^2}}}
	\\ \nonumber
	& + \frac{{{m_s}\left\langle {{g_s}\bar s\sigma Gs} \right\rangle {\alpha _s}}}{\pi })\times {e^{ - s/M_B^2}}ds+ ( - \frac{{\left\langle {g_s^3{G^3}} \right\rangle m_s^2}}{{24{\pi ^2}}} + \frac{4}{9}\left\langle {{\alpha _s}GG} \right\rangle {m_s}\left\langle {\bar ss} \right\rangle
	\\ \nonumber
	&- \frac{{\left\langle {{\alpha _s}GG} \right\rangle {m_s}\left\langle {\bar ss} \right\rangle {\alpha _s}}}{{12\pi }} - \frac{{16\pi m_s^2{{\left\langle {\bar ss} \right\rangle }^2}{\alpha _s}}}{9} - \frac{4}{3}\pi \left\langle {\bar ss} \right\rangle \left\langle {{g_s}\bar s\sigma Gs} \right\rangle {\alpha _s}) \, ,
	\\
	\widetilde{\Pi}^{\alpha\beta}_{1^{-+}} {\left( {M_B^2,{s_0}} \right)} &= \int_{4m_s^2}^{{s_0}} {(\frac{{{s^3}{\alpha _s}}}{{15{\pi ^3}}}}  - \frac{{4m_s^2{s^2}{\alpha _s}}}{{3{\pi ^3}}} + s(\frac{{19\left\langle {{\alpha _s}GG} \right\rangle {\alpha _s}}}{{216{\pi ^3}}} + \frac{{32{m_s}\left\langle {\bar ss} \right\rangle {\alpha _s}}}{{9\pi }}) + \frac{{2\left\langle {{\alpha _s}GG} \right\rangle m_s^2}}{{3{\pi ^2}}}
	\\ \nonumber
	&- \frac{{4{m_s}\left\langle {{g_s}\bar s\sigma Gs} \right\rangle {\alpha _s}}}{\pi })\times {e^{ - s/M_B^2}}ds+ ( - \frac{{\left\langle {g_s^3{G^3}} \right\rangle m_s^2}}{{2{\pi ^2}}} - \frac{{16}}{9}\left\langle {{\alpha _s}GG} \right\rangle {m_s}\left\langle {\bar ss} \right\rangle
	\\ \nonumber
	&  - \frac{{\left\langle {{\alpha _s}GG} \right\rangle {m_s}\left\langle {\bar ss} \right\rangle {\alpha _s}}}{{3\pi }}- \frac{{64\pi m_s^2{{\left\langle {\bar ss} \right\rangle }^2}{\alpha _s}}}{9} + \frac{{16}}{3}\pi \left\langle {\bar ss} \right\rangle \left\langle {{g_s}\bar s\sigma Gs} \right\rangle {\alpha _s}) \, ,
	\\
	\Pi^{\alpha\beta}_{1^{-+}} {\left( {M_B^2,{s_0}} \right)}&= \int_{4m_s^2}^{{s_0}} {(\frac{{{s^3}{\alpha _s}}}{{60{\pi ^3}}}}  - \frac{{m_s^2{s^2}{\alpha _s}}}{{3{\pi ^3}}} + s(\frac{{19\left\langle {{\alpha _s}GG} \right\rangle {\alpha _s}}}{{864{\pi ^3}}} + \frac{{8{m_s}\left\langle {\bar ss} \right\rangle {\alpha _s}}}{{9\pi }}) + \frac{{\left\langle {{\alpha _s}GG} \right\rangle m_s^2}}{{6{\pi ^2}}}
	\\ \nonumber
	&  - \frac{{{m_s}\left\langle {{g_s}\bar s\sigma Gs} \right\rangle {\alpha _s}}}{\pi })\times {e^{ - s/M_B^2}}ds+ ( - \frac{{\left\langle {g_s^3{G^3}} \right\rangle m_s^2}}{{8{\pi ^2}}} - \frac{4}{9}\left\langle {{\alpha _s}GG} \right\rangle {m_s}\left\langle {\bar ss} \right\rangle
	\\ \nonumber
	&- \frac{{\left\langle {{\alpha _s}GG} \right\rangle {m_s}\left\langle {\bar ss} \right\rangle {\alpha _s}}}{{12\pi }}- \frac{{16\pi m_s^2{{\left\langle {\bar ss} \right\rangle }^2}{\alpha _s}}}{9} + \frac{4}{3}\pi \left\langle {\bar ss} \right\rangle \left\langle {{g_s}\bar s\sigma Gs} \right\rangle {\alpha _s}) \, ,
	\\
	\widetilde{\Pi}^{\alpha\beta}_{1^{++}} {\left( {M_B^2,{s_0}} \right)} &= \int_{4m_s^2}^{{s_0}} {(\frac{{{s^3}{\alpha _s}}}{{15{\pi ^3}}}}  - \frac{{4m_s^2{s^2}{\alpha _s}}}{{3{\pi ^3}}} + s(\frac{{19\left\langle {{\alpha _s}GG} \right\rangle {\alpha _s}}}{{216{\pi ^3}}} + \frac{{32{m_s}\left\langle {\bar ss} \right\rangle {\alpha _s}}}{{9\pi }}) - \frac{{2\left\langle {{\alpha _s}GG} \right\rangle m_s^2}}{{3{\pi ^2}}}
	\\ \nonumber
	& + \frac{{4{m_s}\left\langle {{g_s}\bar s\sigma Gs} \right\rangle {\alpha _s}}}{\pi })\times {e^{ - s/M_B^2}}ds+ ( - \frac{{\left\langle {g_s^3{G^3}} \right\rangle m_s^2}}{{6{\pi ^2}}} + \frac{{16}}{9}\left\langle {{\alpha _s}GG} \right\rangle {m_s}\left\langle {\bar ss} \right\rangle
	\\ \nonumber
    &- \frac{{\left\langle {{\alpha _s}GG} \right\rangle {m_s}\left\langle {\bar ss} \right\rangle {\alpha _s}}}{{3\pi }}- \frac{{64\pi m_s^2{{\left\langle {\bar ss} \right\rangle }^2}{\alpha _s}}}{9} - \frac{{16}}{3}\pi \left\langle {\bar ss} \right\rangle \left\langle {{g_s}\bar s\sigma Gs} \right\rangle {\alpha _s}) \, ,
	\\
	\Pi^{\alpha_1\beta_1,\alpha_2\beta_2}_{2^{++}} {\left( {M_B^2,{s_0}} \right)} &= \int_{4m_s^2}^{{s_0}} {(\frac{{2{s^3}{\alpha _s}}}{{15{\pi ^3}}}}  - \frac{{22m_s^2{s^2}{\alpha _s}}}{{5{\pi ^3}}} + s(\frac{{\left\langle {{\alpha _s}GG} \right\rangle }}{{3{\pi ^2}}} - \frac{{89\left\langle {{\alpha _s}GG} \right\rangle {\alpha _s}}}{{144{\pi ^3}}} + \frac{{80{m_s}\left\langle {\bar ss} \right\rangle {\alpha _s}}}{{3\pi }})
	\\ \nonumber
	 &- \frac{{2\left\langle {{\alpha _s}GG} \right\rangle m_s^2}}{{{\pi ^2}}} - \frac{{12{m_s}\left\langle {{g_s}\bar s\sigma Gs} \right\rangle {\alpha _s}}}{\pi }) \times {e^{ - s/M_B^2}}ds+ ( - \frac{{{{\left\langle {{\alpha _s}GG} \right\rangle }^2}}}{{288{\pi ^2}}} - \frac{{3\left\langle {g_s^3{G^3}} \right\rangle m_s^2}}{{{\pi ^2}}}
	\\ \nonumber
	&+ \frac{8}{3}\left\langle {{\alpha _s}GG} \right\rangle {m_s}\left\langle {\bar ss} \right\rangle  + \frac{{2\left\langle {{\alpha _s}GG} \right\rangle {m_s}\left\langle {\bar ss} \right\rangle {\alpha _s}}}{\pi } - \frac{{176}}{3}\pi \left\langle {\bar ss} \right\rangle \left\langle {{g_s}\bar s\sigma Gs} \right\rangle {\alpha _s}) \, ,
	\\
	\widetilde{\Pi}^{\alpha_1\beta_1,\alpha_2\beta_2}_{2^{-+}} {\left( {M_B^2,{s_0}} \right)} &= \int_{4m_s^2}^{{s_0}} {(\frac{{8{s^3}{\alpha _s}}}{{15{\pi ^3}}}}  - \frac{{88m_s^2{s^2}{\alpha _s}}}{{5{\pi ^3}}} + s( - \frac{{4\left\langle {{\alpha _s}GG} \right\rangle }}{{3{\pi ^2}}} - \frac{{17\left\langle {{\alpha _s}GG} \right\rangle {\alpha _s}}}{{36{\pi ^3}}}
	\\ \nonumber
	&+ \frac{{320{m_s}\left\langle {\bar ss} \right\rangle {\alpha _s}}}{{3\pi }}) + \frac{{8\left\langle {{\alpha _s}GG} \right\rangle m_s^2}}{{{\pi ^2}}} + \frac{{48{m_s}\left\langle {{g_s}\bar s\sigma Gs} \right\rangle {\alpha _s}}}{\pi }) \times {e^{ - s/M_B^2}}ds
	\\ \nonumber
	&+ (\frac{{{{\left\langle {{\alpha _s}GG} \right\rangle }^2}}}{{72{\pi ^2}}} + \frac{{8\left\langle {g_s^3{G^3}} \right\rangle m_s^2}}{{{\pi ^2}}} - \frac{{32}}{3}\left\langle {{\alpha _s}GG} \right\rangle {m_s}\left\langle {\bar ss} \right\rangle  - \frac{{1088}}{3}\pi \left\langle {\bar ss} \right\rangle \left\langle {{g_s}\bar s\sigma Gs} \right\rangle {\alpha _s}) \, ,
	\\
	\Pi^{\alpha_1\beta_1,\alpha_2\beta_2}_{2^{-+}} {\left( {M_B^2,{s_0}} \right)}  &= \int_{4m_s^2}^{{s_0}} {(\frac{{2{s^3}{\alpha _s}}}{{15{\pi ^3}}}}  - \frac{{2m_s^2{s^2}{\alpha _s}}}{{5{\pi ^3}}} + s(\frac{{\left\langle {{\alpha _s}GG} \right\rangle }}{{3{\pi ^2}}} - \frac{{89\left\langle {{\alpha _s}GG} \right\rangle {\alpha _s}}}{{144{\pi ^3}}}- \frac{{16{m_s}\left\langle {\bar ss} \right\rangle {\alpha _s}}}{\pi })
    \\ \nonumber
	& + \frac{{2\left\langle {{\alpha _s}GG} \right\rangle m_s^2}}{{{\pi ^2}}} + \frac{{12{m_s}\left\langle {{g_s}\bar s\sigma Gs} \right\rangle {\alpha _s}}}{\pi }) \times {e^{ - s/M_B^2}}ds+( - \frac{{{{\left\langle {{\alpha _s}GG} \right\rangle }^2}}}{{288{\pi ^2}}} - \frac{{3\left\langle {g_s^3{G^3}} \right\rangle m_s^2}}{{{\pi ^2}}}
	\\ \nonumber
	&- 8\left\langle {{\alpha _s}GG} \right\rangle {m_s}\left\langle {\bar ss} \right\rangle  + \frac{{2\left\langle {{\alpha _s}GG} \right\rangle {m_s}\left\langle {\bar ss} \right\rangle {\alpha _s}}}{\pi } + \frac{{176}}{3}\pi \left\langle {\bar ss} \right\rangle \left\langle {{g_s}\bar s\sigma Gs} \right\rangle {\alpha _s}) \, ,
	\\
	\widetilde{\Pi}^{\alpha_1\beta_1,\alpha_2\beta_2}_{2^{++}} {\left( {M_B^2,{s_0}} \right)} &= \int_{4m_s^2}^{{s_0}} {(\frac{{8{s^3}{\alpha _s}}}{{15{\pi ^3}}}}  - \frac{{8m_s^2{s^2}{\alpha _s}}}{{5{\pi ^3}}} + s( - \frac{{4\left\langle {{\alpha _s}GG} \right\rangle }}{{3{\pi ^2}}} - \frac{{17\left\langle {{\alpha _s}GG} \right\rangle {\alpha _s}}}{{36{\pi ^3}}}
	\\ \nonumber
	 &- \frac{{64{m_s}\left\langle {\bar ss} \right\rangle {\alpha _s}}}{\pi }) - \frac{{8\left\langle {{\alpha _s}GG} \right\rangle m_s^2}}{{{\pi ^2}}} - \frac{{48{m_s}\left\langle {{g_s}\bar s\sigma Gs} \right\rangle {\alpha _s}}}{\pi }) \times {e^{ - s/M_B^2}}ds
	\\ \nonumber
	&+ (\frac{{{{\left\langle {{\alpha _s}GG} \right\rangle }^2}}}{{72{\pi ^2}}} + \frac{{16\left\langle {g_s^3{G^3}} \right\rangle m_s^2}}{{{\pi ^2}}} + 32\left\langle {{\alpha _s}GG} \right\rangle {m_s}\left\langle {\bar ss} \right\rangle  + \frac{{1088}}{3}\pi \left\langle {\bar ss} \right\rangle \left\langle {{g_s}\bar s\sigma Gs} \right\rangle {\alpha _s}) \, .
\end{align}
\end{widetext}

\bibliographystyle{elsarticle-num}
\bibliography{ref}

\end{document}